\documentclass[twocolumn,showpacs,preprintnumbers,amsmath,amssymb]{revtex4}

\usepackage{graphicx}
\usepackage{dcolumn}
\usepackage{bm}
\usepackage{epsfig}
\newcommand{\ce}{{\rm ce}}
\newcommand{\se}{{\rm se}}
\newcommand{\be}{\begin{equation}}
\newcommand{\ee}{\end{equation}}
\newcommand{\Om}{\Omega}
\newcommand{\omr}{\omega_r}
\newcommand{\om}{\omega}
\newcommand{\Omeff}{\Omega_{\rm eff}}
\newcommand{\erp}{{\rm erp}}

\begin{document}


\title{Atom-wave diffraction between the Raman-Nath and the Bragg regime: \\
Effective Rabi frequency, losses, and phase shifts.}
\author{Holger M\"uller}\email{holgerm@stanford.edu}
\author{Sheng-wey Chiow}
\author{Steven Chu}
\altaffiliation{Lawrence Berkeley National Laboratory and
Department of Physics, University of California, Berkeley, 1
Cyclotron Road, Berkeley, CA 94720.}
\affiliation{Physics
Department, Stanford University, Stanford, CA94305.}
\date{\today}

\begin{abstract}
We present an analytic theory of the diffraction of (matter) waves
by a lattice in the ``quasi-Bragg" regime, by which we mean the
transition region between the long-interaction Bragg and
``channelling" regimes and the short-interaction Raman-Nath
regime. The Schr\"odinger equation is solved by {\em adiabatic
expansion}, using the conventional adiabatic approximation as a
starting point, and re-inserting the result into the Schr\"odinger
equation to yield a second order correction. Closed expressions
for arbitrary pulse shapes and diffraction orders are obtained and
the losses of the population to output states otherwise forbidden
by the Bragg condition are derived. We consider the phase shift
due to couplings of the desired output to these states that
depends on the interaction strength and duration and show how
these can be kept negligible by a choice of smooth (e.g.,
Gaussian) envelope functions even in situations that substantially
violate the adiabaticity condition. We also give an efficient
method for calculating the effective Rabi frequency (which is
related to the eigenvalues of Mathieu functions) in the
quasi-Bragg regime.
\end{abstract}

\maketitle


\section{Introduction}

\subsection{Background}

Diffraction by a point scatters light or matter waves into all
directions. A two-dimensional grating produces a few diffraction
orders at those angles where the scatter from all of the grating
adds coherently. Bragg diffraction by an infinite
three-dimensional lattice can produce a single diffraction order,
which happens when the scatter from all layers adds
constructively, as described by the Bragg condition. When this
happens for a higher scattering order (``high-order Bragg
diffraction") virtually all incident radiation can be scattered
into this high order, in contrast to the two-dimensional case. By
{\em quasi-Bragg diffraction}, we refer to the intermediate regime
where the infinite lattice assumption is no longer valid but
approximately true. Using the nomenclature of, e.g.,
\cite{Keller}, this regime is the transition between the
short-interaction Raman-Nath regime and the long-interaction Bragg
(weak potential) and ``channelling" (strong potential) regimes. In
this region, the Bragg condition softens and there may be
significant scattering into other than the desired orders.
Moreover, couplings between the nonzero diffraction orders may
lead to phase shifts of the diffracted waves \cite{Buchner}, which
is undesirable in many applications.

In this work, we present an analytic treatment of such quasi Bragg
scattering. We will find that by prudent choice of the scattering
potential and its envelope function, behavior very similar to
Bragg scattering, in particular very low losses and phase shifts,
can be obtained for scatterers that substantially violate the
assumptions of the simplified theory.

Bragg diffraction famously provides us with the basic knowledge of
the structure of crystals, including proteins. It is also
important for many technical applications, like acousto-optic
modulators (AOMs) \cite{Hobbs}, distributed Bragg reflectors (DBR)
in diode \cite{Voges} and fiber lasers as well as photonic bandgap
crystals \cite{vanDriel}. Moreover, Bragg diffraction is a basic
method for making surface acoustic wave (SAW) filters in radio
frequency technology. In atomic physics, Bragg diffraction is a
special case of the Kapitza-Dirac effect
\cite{Kapitza,Altshule,Martin,Berman,Batelaan}.

Bragg scattering is used as a tool for experiments with
Bose-Einstein condensates (BECs) \cite{Gupta,Gupta2,Morsch}. For
example, Kozuma {\em et al.} \cite{Kozuma} have shown
experimentally that thirteen subsequent first-order Bragg
diffractions of a BEC can still have good efficiency. More exotic
applications include the generation of a collective frictional
force in an ensemble of atoms enclosed in a cavity, due to Bragg
scattering of a pump light an a self-organized atomic density
grating \cite{Black}, much in the same way as stimulated Brillouin
scattering by self-organized acoustic waves in optical fibers
\cite{Voges}.

Moreover, Bragg diffraction can act as a beam splitter for matter
waves
\cite{Martin,Altshule,Meystre,Gupta,Gupta2,Bernet,Wu,Borde,Malinovsky}.
The highest order diffraction so far achieved with matter waves
seems to be by Koolen {\em et al.} \cite{Koolen}, who obtained up
to eighth-order Bragg diffraction. Atom interferometers based on
Bragg diffraction include the one by Giltner {\em et al.}
\cite{Giltner,Giltner2}, who built a Mach-Zehnder atom
interferometer using up to third order diffraction. Miller {\em et
al.} \cite{Miller} achieved high contrast in a two-pulse geometry
with first-order diffraction and a sufficiently short time between
pulses. Torii {\em et al.} \cite{Torii} have used first order
Bragg diffraction in a Mach-Zehnder geometry with a Bose-Einstein
condensate. In addition, Rasel {\em et al.} \cite{Rasel} have
built a Mach-Zehnder atomic-beam interferometer based on
Raman-Nath scattering.

More generally, atom interferometers can be used for measurements
of atomic properties \cite{Ekstrom,Miffre,Schmiedmayer}, the local
gravitational acceleration \cite{Peters}, the gravity gradient,
Newton's gravitational constant \cite{Stuhler}, tests of the
equivalence principle \cite{Jason} and the fine-structure constant
via $h/m$ \cite{Wicht,Biraben,Paris}. For planned experiments in
space, see \cite{Coq,Yu}.

While not all the atom interferometers just cited use Bragg
diffraction, high-order Bragg diffraction offers several
interesting possibilities for atom interferometers: (i) it makes
the atom interact with $2n$ photons at once, which may increase
the sensitivity of the interferometer by a factor of $n^2$
\cite{Paris} relative to the $2-$ photon transitions. (We note
here that other possibilities exist for using high-order
transitions in atom interferometers, like applying multiple
low-order pulses \cite{Giurk}, operation in the Raman-Nath regime
\cite{Dubetsky} or the magneto-optical beam splitter \cite{Pfau})
(ii) Since Bragg diffraction theoretically allows coherent
momentum transfer with an efficiency close to one, it allows the
insertion of many $\pi$-pulses for additional momentum transfer,
which increases the signal in photon recoil measurements
\cite{Paris} (up to $N=30$ $\pi$-pulses based on two-photon
adiabatic transfer were used in \cite{Wicht}, transferring $60$
photon momenta; if these had been 5-th order Bragg pulses, they
would have transferred 300). (iii) If losses can be neglected,
Bragg diffraction is basically a transition in a 2-level system.
Thus, many of the techniques developed for standard beam splitters
based on Raman transitions can be taken over. For example, several
beam splitters addressing different velocity groups respectively
can be performed simultaneously \cite{Paris,PLL}.

\subsection{Overview of the existing theory}

A summary of the material that is the basis of this work can be
found in the textbook by Meystre \cite{Meystre}.

A lot of attention has been paid on the theory of the
long-interaction time (channelling or Bragg) regimes on the one
hand and the Raman-Nath regime on the other hand. Keller {\em et
al.} \cite{Keller} give a brief account of the most important
results. They have been derived using various formalisms: Berman
and Bian \cite{Berman} use a pump-probe spectroscopy picture,
focussing on applications as beam splitters in atom
interferometers. The phase-shift of the diffraction process has
been studied by B\"uchner {\em et al.} \cite{Buchner}, limited to
first and second order diffraction. Giltner {\em et al.}
\cite{Giltner,Giltner2} have reported an atom interferometer based
on Bragg diffraction of up to third order and give the effective
Rabi frequency in the long-interaction regime, our Eq.
(\ref{Omeff}). A similar derivation was given by Gupta {\em et
al.} \cite{Gupta,Gupta2}.

While most of this work (as well as ours) is concerned with
on-resonance transitions, D\"urr and Rempe \cite{Duerr99} have
considered the acceptance angle (i.e., linewidth) of diffraction.
They restrict attention to the case of square-envelope pulses. Wu
et al. \cite{Wu} give a numerical study of the latter, again for
the case of (a pair of) square pulses. Stenger {\em et al.}
\cite{Stenger} describe the line shape of Bragg diffraction with
Bose-Einstein condensates within the Bragg regime.

More general types of diffraction have been studied, such as using
chirped laser frequencies \cite{Malinovsky} in the adiabatic
regime where the chirp is sufficiently slow. This regime is
similar to Bloch oscillations \cite{Peik}. Band {\em et al.}
\cite{Band} have considered the loss due to atom-atom
interactions, which is relevant in experiments with Bose-Einstein
condensates (BECs). Blackie and Ballagh \cite{Blackie} explore the
use of Bragg diffraction in probing vortices in BECs.

On the other hand, the picture is still incomplete in the
quasi-Bragg regime we are concerned with. D\"urr and Rempe have
analytically calculated corrections to the effective Rabi
frequency for large potential depth \cite{Duerr} in the case of
second-order scattering with a square envelope function.
Champenois {\em et al.} \cite{Champenois} consider both the
Raman-Nath and the Bragg regime using a Bloch-state approach; for
the Bragg regime, however, their treatment is restricted to the
first diffraction order and square envelopes. Bord\'{e} and
L\"ammerzahl \cite{Borde} give and exhaustive treatment of
square-envelope scattering that is based on matching the boundary
conditions at the beginning and end of the pulse. This method,
unfortunately, cannot readily be generalized to smooth envelope
functions.

The Mathieu equation formalism is a powerful tool for studying
diffraction with arbitrary potential depth and interaction times,
as demonstrated by Horne, Jex, and Zeilinger \cite{Horne}.
However, like the work of Bord\'{e} and L\"ammerzahl, the Mathieu
equation formalism assumes constant envelopes and cannot readily
be generalized to smooth envelope functions. As we shall see (and
has already been pointed out, see, e.g., \cite{Keller}), smooth
envelope functions are of particular interest because they allow
high-efficiency scattering into a single order even in the
quasi-Bragg regime. The mathematical properties of the Mathieu
functions themselves have been explored by many workers. Of
relevance for this work is the power-series expansion of Mathieu
functions reported by Kokkorakis and Roumelotis \cite{Kokkorakis}.

\subsection{Motivation}

In the experimental applications, it is often desirable to make
the interaction time as short as compatible with certain
requirements on the efficiency and parasitic phase shifts. This
can be achieved by operating in the quasi-Bragg regime. For
example, in atomic physics, long interaction times increase losses
due to single-photon excitation and also systematic effects in
atom interferometers. For fifth order Bragg scattering of cesium
atoms in a standing light wave, satisfying the adiabaticity
criterion requires interaction times $\gg 0.4$\,s, see Sec.
\ref{adiabaticbragg}. This exceeds the time available in
experiments under free-fall conditions, and would give rise to
huge losses by single-photon excitation. Moreover, the very sharp
Bragg condition in the pure Bragg regime means that scattering
happens only if the incident waves are within a very narrow of the
velocity distribution of a thermal sample, which may mean that a
large fraction will not be scattered at all. However, operation in
the quasi-Bragg regime requires a theoretical calculation of the
losses and phase-shifts encountered, and strategies to minimize
them. This is best done by analytic equations for these
parameters, which allow one to easily see which parameters have to
take which values.

Unfortunately, in this regime the population of the diffraction
orders is particularly difficult to calculate \cite{Keller} and so
the existing analytic theory of the quasi-Bragg regime is
restricted to square envelope functions or low orders, or both.
Moreover, even then, using the known formalisms it is hard to
obtain power-series approximations of parameters like the
effective Rabi frequency in terms of the interaction strength.

The aim of this paper is to present an analytic theory of the
quasi-Bragg regime that allows the treatment of arbitrary
scattering orders and envelope functions of the scattering
potential. To do so, we develop a systematic way of obtaining more
and more accurate solutions of the Schr\"odinger equation that
starts from the usual adiabatic approximation. This allows us to
calculate the population of the diffraction orders, including
losses to unwanted outputs and phase shifts. We can thus specify
the minimum interaction time and the maximum interaction strength
that yield losses which are below a given level, and consider the
influence of the pulse shape. It will turn out that efficient
scattering can be maintained with interaction times that
substantially violate the adiabaticity criterion, in agreement
with experiments \cite{Keller}. For example, we show that fifth
order scattering of Cs atoms still has negligible losses and
phase-shifts for interaction times on the order of 10\,$\,\mu$s if
the pulse shape of the light is appropriately chosen. This minimum
interaction time even decreases for higher scattering order. We
will restrict attention to on-resonant Bragg diffraction,
neglecting any initial velocity spread that the atoms may have.
This can certainly be a good assumption for experiments using
Bose-Einstein condensed atoms, but also for a much wider class of
experiments: The minimum interaction time that will still lead to
low losses will turn out to be roughly given by $1/(n \omr)$,
where $\omr$ is the recoil frequency and $n$ the Bragg diffraction
order, see Eq. (\ref{Gaussianlimit2}). Such short pulses have a
Fourier linewidth that is on the order of $n\omr$. This means that
a velocity spread on the order of the recoil velocity cannot be
resolved, even though the velocity selectivity increases $\sim n$
due to the multiple photons scattered. The preparation of atoms
having a velocity spread of 1/100-1/10 of the recoil velocity is
standard practice in atomic fountains.

While our theory will be stated in the language of atomic physics,
with an eye on applications in atom interferometry, it can be
adapted to Bragg diffraction in all fields of physics.

\subsection{Outline}

This paper is organized as follows: In Sec. \ref{Problem}, we
describe our basic Hamiltonian and the conventional theory for the
Bragg and the Raman-Nath regime. In Sec. \ref{Mathieusec}, an
exact solution for rectangular envelope functions will be
presented that uses Mathieu functions. In Sec.
\ref{Rabifreqmethod}, we find a general form for the corrections
of the effective Rabi frequency which is valid for arbitrary
scattering orders. In Sec. \ref{Highorderadiabat} we present our
method for calculating the phase shifts and losses, adiabatic
expansion. In Sections \ref{squarepuls} and \ref{Gaussianpuls}, we
consider square and Gaussian envelope functions and give a
practical example of high-order Bragg scattering of Cs atoms.

\subsection{Problem}
\label{Problem}

In the remainder of this Sec. I, we define the basic problem and
review the basic theory of the adiabatic and the Raman-Nath case
(as described, e.g., in \cite{Meystre}) and of the Mathieu
equation approach. This is to define the notation and for the
reader's convenience. The reader already familiar with this may
want to proceed to the following sections, which describe the new
results of this paper.

Consider scattering of an atom of mass $M$ by a standing wave of
light along the $z$ direction. Ignoring effects of spontaneous
emission, the Hamiltonian describing the interaction of the atoms
with the standing wave having a wavenumber $k$ is (in a frame
rotating at the laser frequency $\om$)
\begin{equation}
H=\frac{p^2}{2M}-\hbar \delta |e\rangle\langle
e|+\hbar\Om_0(t)\cos(kz)\left(|e\rangle\langle g|+H.c.\right)\,,
\end{equation}
where $[z,p]=i\hbar$ and $\delta$ is the detuning. The Rabi
frequency $\Om_0$ may in general be time-dependent. For the
purpose of this introduction, we will assume it to be constant. In
the later sections of this paper, we shall be interested in the
effects of different pulse shapes, however. Substituting
\begin{equation}
|\psi(t)\rangle=e(z,t)|e\rangle+g(z,t)|g\rangle
\end{equation}
into the Schr\"odinger equation yields the coupled differential
equations
\begin{eqnarray}
i\hbar\dot
e(z,t)&=&\frac{p^2}{2M}e(z,t)+\hbar\Om_0\cos(kz)g(z,t)-\hbar\delta
e(z,t)\,,\nonumber \\
i\hbar\dot
g(z,t)&=&\frac{p^2}{2M}g(z,t)+\hbar\Om_0\cos(kz)e(z,t)\,,
\end{eqnarray}
where the dot denotes the time derivative. For $\delta$ large
compared to the linewidth of the excited state (and thus also
$\delta\gg \Om_0, \omr$) and the atoms initially in the ground
state, we can adiabatically eliminate the excited state:
\begin{equation}
\label{Mathieu} i\hbar\dot
g(z,t)=-\frac{\hbar^2}{2M}\frac{\partial^2g(z,t)}{\partial
z^2}+\frac{\hbar\Om_0^2}{\delta}\cos^2(kz)g(z,t)\,.
\end{equation}
This equation with its periodic potential is invariant under a
translation by an integer multiple of $k^{-1}$. Applying the Bloch
theorem, we can look for solutions having constant quasi-momentum;
in particular, we can restrict attention to the case of vanishing
quasi-momentum. For constant $\Om_0$, this is a Mathieu equation
for which exact solutions are known; this formalism will be
described in Sec. \ref{Mathieusec} and developed further in Sec
\ref{Rabifreqmethod}. If we let
\begin{equation}
g(z,t)=\sum_{m=-\infty}^\infty g_m(t)e^{imkz}
\end{equation}
and use $\cos^2(kz)=1/2+1/4(e^{2ikz}+e^{-2ikz})$, we obtain
\begin{eqnarray}
i\hbar\sum_{m=-\infty}^\infty \dot
g_m(t)e^{imkz}=\hbar\sum_{m=-\infty}^\infty [(\omr m^2+\Om)g_m
\nonumber
\\ +(\Om/2)(g_{m+2}+g_{m-2})]e^{imkz}
\end{eqnarray}
where we have introduced the two-photon Rabi frequency
\begin{equation}
\Om=\frac{\Om_0^2}{2\delta}
\end{equation}
and the recoil frequency
\begin{equation}
\omr=\frac{\hbar k^2}{2M}\,.
\end{equation}
This can only hold if for all $m$
\begin{equation}\label{adiabatelim}
i\hbar \dot g_m=\hbar(\omr
m^2+\Om)g_m+\hbar(\Om/2)(g_{m+2}+g_{m-2})\,.
\end{equation}
[Since this equation couples only odd or even momentum states,
respectively, we can and will look for solutions that have either
the odd or even terms zero. In view of this the use of both even
and odd indices may seem unnecessary, but will have advantages
when we consider Bragg diffraction.] The theoretical description
of Bragg diffraction is relatively simple in the short-interaction
limit (the Raman-Nath regime) and in the case of an infinite
scatterer, the Bragg regime.

\subsubsection{Raman-Nath Regime}

The Raman-Nath regime is defined as the case of very short
interaction time, so that the kinetic energy term is negligible
against the resulting energy uncertainty. Equation
(\ref{adiabatelim}) reduces to
\begin{equation}
i\hbar \dot g_m=\hbar(\Om/2)(g_{m+2}+g_{m-2})\,,
\end{equation}
which we have simplified by shifting the energy scale by
$-\hbar\Om$. Since these equations only couple states which differ
by an even multiple of the momentum $\hbar k$, we can restrict
attention to even indices $2m$. They can be satisfied by Bessel
functions:
\begin{equation}
g_{2m}=(-i)^mJ_m(\Om t)\,.
\end{equation}
At $t=0$, this solution has all atoms in the zero momentum state
$g_0=1, g_{2m\neq 0}=0$. For $t\neq 0$, the probability to find
the atom to have a transverse momentum $2m\hbar k$ is
$P_{2m}(t)=J_m^2(\Om t)$. The Raman-Nath approximation holds
provided that $t\ll 1/\sqrt{2\Om \omr}$ (because then a high
energy uncertainty justifies our neglect of the kinetic energy).
Clearly, the transfer efficiency $P_{2m}$ for any particular $m$
is limited. For example, the maximum probability to find the atom
in the ground state after scattering 2 photons is approximately
0.34.

\subsubsection{Bragg Regime}\label{adiabaticbragg}

For the Bragg regime, we take into account the kinetic energy term
and work in configuration space. We now assume initial conditions
$g_{-n}=1$ and $g_m=0$ for $m\neq -n$. To simplify, we subtract a
constant offset $n^2\hbar\om_r+\hbar\Om$ from the energy scale.
Eq. (\ref{adiabatelim}) now reads
\begin{eqnarray}\label{momentumeqs}
& \vdots & \nonumber \\
i\hbar \dot g_{-n-2}&=&(4+4n)\hbar\om_rg_{-n-2}+\frac12
\hbar\Om(g_{-n}+g_{-n-4}) \nonumber \\
i\hbar \dot g_{-n}&=&\frac
12 \hbar\Om(g_{-n+2}+g_{-n-2}) \nonumber \\
i\hbar \dot
g_{-n+2}&=&(4-4n)\hbar\om_rg_{-n+2}+\frac12
\hbar\Om(g_{-n+4}+g_{-n}) \nonumber
\\ & \vdots &  \nonumber \\
i\hbar \dot g_{-n+2k}&=&4k(k-n)\hbar\omr g_{-n+2k}\nonumber
\\ && +\frac12
\hbar\Om(g_{-n+2k+2}+g_{-n+2k-2}) \nonumber  \\
& \vdots &
\nonumber \\ i\hbar \dot
g_{n-2}&=&(4-4n)\hbar\om_rg_{n-2}+\frac12 \hbar\Om(g_{n}+g_{n-4}) \nonumber \\
i\hbar g_n&=&\frac 12 \hbar\Om(g_{n+2}+g_{n-2}) \nonumber \\
i\hbar \dot g_{n+2}&=&(4+4n)\hbar\om_rg_{n+2}+\frac12
\hbar\Om(g_{n+4}+g_{n}) \nonumber
\\ & \vdots & \,.
\end{eqnarray}
Energy conservation will favor transitions from $-n \rightarrow
n$, if the processes are sufficiently slow. This is the result of
the adiabatic elimination of the intermediate states ($k\neq 0$
and $k\neq n$): If
\begin{equation}
|4k^2-4nk|\hbar\om_r \gg \hbar\Om,
\end{equation}
for all $0<k<n$, we can assume that the $k$th equation is always
in equilibrium with $\dot g_{-n+2k}\approx 0$. Then, for example,
\begin{equation} g_{-n+2}=-\frac18
\hbar\Om\frac{1}{(nk-k^2)\hbar\om_r}g_{-n}\, .
\end{equation} Relations like this can be used to successively eliminate all $n-1$
intermediate states. With \begin{equation}\label{Omeff}
\Omeff=\frac{\Om^n}{(8\om_r)^{n-1}}\prod_{k=1}^{n-1}\frac{1}{nk-k^2}
=\frac{\Om^n}{(8\om_r)^{n-1}}\frac{1}{[(n-1)!]^2}\,,
\end{equation} we obtain
\begin{eqnarray}\label{adiabaticdiffeq}
i\hbar \dot g_{-n}&=&\frac 12 \hbar\Omeff
 g_n \nonumber \\ i\hbar \dot g_{n}&=&\frac 12
\hbar\Omeff g_{-n}
\end{eqnarray}
(where we have removed a constant light shift term). This can be
readily solved:
\begin{equation}
\label{adiabaticsolution}
g_{-n}(t)=\cos\frac12\Omeff t\,, \quad
g_n(t)=-i\sin\frac12\Omeff t \,.
\end{equation}
For a time-varying $\Om$,
\begin{eqnarray}\label{adiabaticint}
g_{-n}(t)&=&\cos\left(\frac12 \int_{-\infty}^t \Omeff(t')
dt'\right) \,,\nonumber \\
g_n(t)&=& -i\sin\left(\frac 12
\int_{-\infty}^t\Omeff(t')dt'\right) \,.
\end{eqnarray}
This is an exact solution of the adiabatic equations of motion
Eqs. (\ref{adiabaticdiffeq}) for real $\Omeff$, as can be verified
by insertion. If the integral appearing in the trigonometric
functions is equal to $\pi$ (a ``$\pi$-pulse") , all of the
population ends up in the final state; if it is $\pi/2$ (a
``$\pi/2$-pulse"), half of it.

While operation in the Bragg regime is lossless, it requires
relatively long interaction times. In the previous section, we
used the condition $4(1-n)\om_r \gg \Om$, which translates into
\begin{equation}\label{adiabaticityBragg}
\Omeff\ll \frac{8(n-1)^n\omr}{2^n(n-1)!^2}.
\end{equation}
This is $\sim \omr$ for $n\leq 5$, but drops rapidly, e.g.,
$\Omeff\ll 2\times 10^{-4}\omr$ for $n=10$.

For later use, we consider the case of complex
$\Omeff=|\Omeff|e^{i\varphi}$, where $\varphi$ is the argument of
$\Omeff$. Hermiticity then requires us to use $\Omeff^\ast$ in the
second of Eqs. (\ref{adiabaticdiffeq}). The solution of these
equations for constant $\Omega$ is $g_{-n}(t)=\cos\frac12|\Omeff|
t, g_n(t)=-ie^{-i\varphi}\sin\frac12|\Omeff| t$. For time-varying
complex $\Omega$, the generalization of Eqs. (\ref{adiabaticint})
by substituting $|\Omeff|$ and inserting the factor of
$e^{-i\varphi}$ into $g_n(t)$ is an exact solution for constant
$\varphi$, and remains approximately valid as $|\dot \varphi| \ll
|\Omeff|/2$, as can be seen by inserting into Eqs.
(\ref{adiabaticdiffeq}). Thus, this solution holds for varying
complex $\Omeff$ as long as its argument changes adiabatically. In
the following, $\Omeff$ is always understood to be the absolute
value unless otherwise stated.

\subsection{Mathieu equation approach}\label{Mathieusec}

For constant $\Om_0$, we can apply the method of separation of
variables to Eq. (\ref{Mathieu}). We are looking for a solution of
the form
\begin{equation} g(z,t)=g_t(t)g(z) \end{equation} to obtain
\begin{eqnarray}
i g_t'-\om_t g_t & = & 0 \nonumber \\
-\frac{\hbar}{2M} g''+2\Omega\cos^2(kz)g-\om_t g & = & 0
\end{eqnarray}
with $\om_t$ being the separation constant. The second equation is
the Mathieu equation. By using $v=z/k$,
\begin{eqnarray}\label{defq} a &
=&-\frac{\Om}{\om_r}+\frac{\om_t}{\omr}\,, \nonumber
\\ q&=&\frac{\Om}{2\omr}\,,
\end{eqnarray}
it can be brought to the standard form
\begin{equation}
g''(v)+[a-2q\cos(2v)]g(v)=0.
\end{equation}
This is an eigenvalue equation in $a$. The eigenfunctions can be
expressed by Fourier series
\begin{eqnarray}
\se_{2n+1}(v,q)&=&\sum_{r=0}^{\infty}
B_{2r+1}^{(2n+1)}\sin(2r+1)v\,,\nonumber \\ \se_{2n+2}(v,q)&=&
\sum_{r=0}^{\infty} B_{2r+2}^{(2n+2)}\sin(2r+2)v\,,\nonumber \\
\ce_{2n}(v,q)&=&\sum_{r=0}^{\infty} A_{2r}^{(2n)}\cos 2r v \,, \nonumber \\
\ce_{2n+1}(v,q)&=&\sum_{r=0}^{\infty} A_{2r+1}^{(2n+1)}\cos(2r+1)v
\,.
\end{eqnarray}
The eigenvalues associated with the $v-$even functions $\ce$ are
denoted $a_r$, those associated with odd functions $\se$ are
denoted $b_r$ \cite{Abramowitz}. Insertion of these series into
the Mathieu equation allows to determine the eigenvalues and the
Fourier coefficients \cite{Abramowitz,Rhyshik}. This is tedious,
but tables \cite{Abramowitz} and standard numerical routines allow
to find numerical values easily.

In this paragraph, we use units with $\omr=1$. For an atom
initially in a pure momentum state with $p=-n\hbar k$, we express
the wave function $\psi(t=0)$ as a series of Mathieu functions:
\begin{equation}
e^{in_i v}=\sum_m C_{m}\ce_{m}+S_{m}\se_{m}.
\end{equation} This is possible, because the functions $\ce$ and
$\se$ form a complete orthogonal set \cite{Abramowitz}. The
coefficients of the expansion are thus given by the Fourier
coefficients of the Mathieu functions. Once the they are known, we
can write down the amplitude of finding the atom with a momentum
$n_f$ at a later time:
\begin{eqnarray}
\label{Mathieuamplitude}
g_{n_f}=\frac{1}{2\pi} \int_0^{2\pi} e^{-in_fv}\left(
\sum_{m} A_{n_i}^{(m)}\ce_{m}(v,q) e^{i (a_{m}+2q)t}
\right.\nonumber
\\ \left. \vphantom{\sum_n}+iB_{n_i}^{(m)}\se_{m}(v,q)
e^{i (b_{m}+2q) t}\right) dv  \nonumber \\
=\frac{e^{2iqt}}{2}\left( \sum_{m} A_{n_i}^{(m)}A_{n_f}^{(m)}e^{i
a_m t}+B_{n_i}^{(m)}B_{n_f}^{(m)} e^{ib_{m} t}\right)
\end{eqnarray}
(The $e^{-in_fv}$ has a minus in the exponent because this is a
reverse Fourier transform.) While in general this is a very
complicated function of $t$, for low values of $q$ only
$B_{n_i}^{(n_i)}$ and $A_{n_i}^{(n_i)}$ will be large. If we
neglect all others,
\begin{eqnarray}\label{Mathieumean}
|g_{n_f}|^2\approx\frac14\left(|A_{n_i}^{(n_i)}|^4+|B_{n_i}^{(n_i)}|^4\right.
\nonumber
\\ \left. \pm 2|A_{n_i}^{(n_i)}|^2||B_{n_i}^{(n_i)}|^2\cos (a_{n_i}-b_{n_i})
t\right)\,.
\end{eqnarray}
The plus sign is for $n_f=n_i$, the minus sign for $n_f=-n_i$.
Thus, the atoms will oscillate between $n_i$ and $-n_i$ with an
effective Rabi frequency $\Omeff \equiv a_{n_i}-b_{n_i}$.

For an explicit example, let the two photon Rabi frequency be
$\Om=2q=3$ for $0<t<T$ and 0 otherwise. Suppose further that for
$t<0$ the atom is in an initial state
$\psi=e^{-i3v}=\cos(3v)-i\sin(3v)$ having momentum $p=-3\hbar k$.
The coefficients are most easily obtained by numerical calculation
of the Fourier integral.

The amplitude of finding the atom with a momentum $+3$ at a later
time is given by Eq. (\ref{Mathieuamplitude}): \begin{equation}
g_3 =\frac12e^{2iqt}\left( \sum_{n} |A_3^{(2n+1)}|^2e^{i
a_{2n+1}t}+|B_3^{(2n+1)}|^2 e^{ib_{2n+1} t}\right) \end{equation}
For the population in the initial state, we find
\begin{eqnarray}
g_{-3}(t) &=& \frac12e^{2iqt}\left( \sum_{n} |A_3^{(2n+1)}|^2e^{i
a_{2n+1}t}\right. \nonumber \\ && \left. -|B_3^{(2n+1)}|^2
e^{ib_{2n+1} t}\right).
\end{eqnarray}

\begin{figure}
\centering \epsfig{file=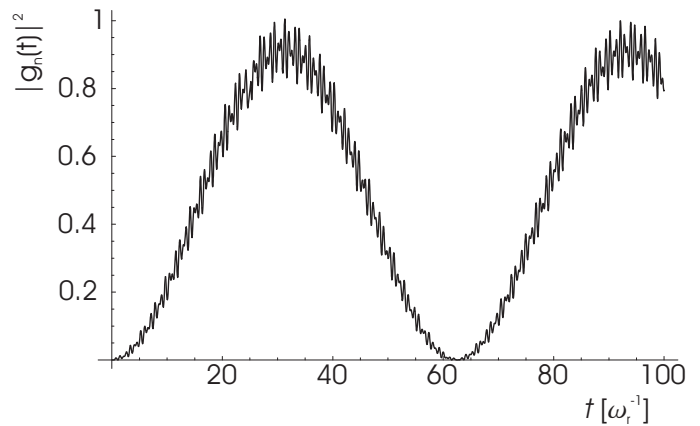,width=0.5\textwidth}
\caption{\label{mathieusolution} Population in the final state
$|g_n(t)|^2$ as obtained from the Mathieu equation approach
plotted versus time $t$ in units of $\omr^{-1}$.}
\end{figure}

\subsubsection{Losses}

The solution (Fig. \ref{mathieusolution}) oscillates quickly
around the mean as given by Eq. (\ref{Mathieumean}). The
frequencies of oscillation are relatively large compared to the
effective Rabi frequency and depend on the pulse amplitudes. Thus,
observing them requires very accurate timing and control over the
pulse amplitudes. Especially, because of interference fringes in
optical setups, it is hard to achieve an amplitude stability
better than about 1\%. Thus, these oscillations may be hard to
observe in practice and the transfer efficiency very close to one
at some of their peaks is not useful in practice. Most of the
time, the population that can practically be achieved will thus be
more close to the mean value as given by Eq. (\ref{Mathieumean}).
In our previous example, this reaches a maximum of 0.914, i.e.,
8.6\% of the population are not transferred to the final state.

\subsubsection{Phase shifts}

\begin{figure}[t]
\centering \epsfig{file=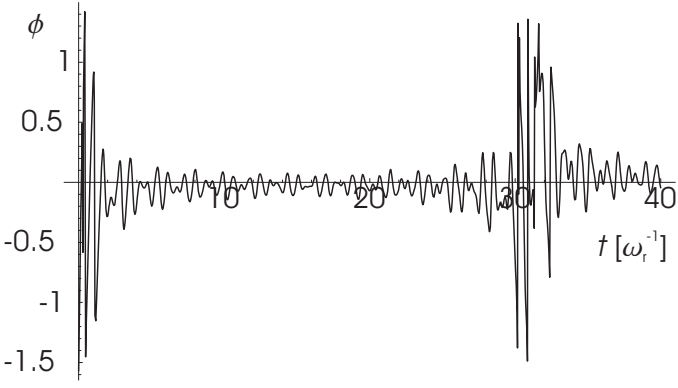, width=0.5\textwidth}
\caption{\label{mathieuphase} Phase $\phi-\pi/2$ of the final
state relative to the initial state plotted versus time $t$ in
units of $\omr^{-1}$.}
\end{figure}

In this picture, the initial and final momentum states have the
same energy. If they were freely propagating (no interactions with
neighbor states), their wave function should thus exhibit the same
phase factor $\exp(iEt/\hbar)$. The interactions, however, cause a
difference of the phase which can be seen by considering the ratio
of the amplitudes
\begin{equation}
\tan \phi= \frac{g_{-3}(t)}{g_3(t)}
\end{equation}
which can be calculated in a straightforward way. The phase is
most conveniently discussed in terms of $\phi-\pi/2$, by
subtracting the phase of $\pi/2$ which is expected in the pure
Bragg regime, compare Eq. (\ref{adiabaticsolution}). As shown in
Fig. \ref{mathieuphase}, this is an oscillating function of time.
Around the time for a $\pi/2$-pulse, the amplitude of the
oscillation of the relative phase changes are given by
$\tan\phi\approx \phi\approx (A_3^{(5)})^2 \approx (B_3^{(5)})^2$.
However, since the phase is an oscillatory function, there are
instances where the phase is larger or vanishes exactly.

For the practical application of this in atom interferometry,
where this relative phase adds to the phase to be measured, two
remarks of caution are appropriate: (i) As mentioned before,
making use of the theoretically exact vanishing of the phase shift
at particular times requires very accurate timing and control over
the pulse amplitudes. (ii) In certain interferometer geometries,
equal parasitic phases of subsequent beam splitters cancel out.
However, since this depends sensitively on very small amplitude
changes of the pulses (that would affect the times of the
zero-crossings of the wiggles in Fig. \ref{mathieuphase}), the
cancellation is impaired. Thus, in practice, it may be impossible
to rely on this exact vanishing or cancellation.

\section{Efficient method for calculating the effective Rabi
frequency} \label{Rabifreqmethod}

The adiabatic elimination process yields a simple equation for the
effective Rabi frequency $\Omeff$. However, it is inappropriate in
the quasi-Bragg regime. Higher order corrections, that tend to
reduce $\Omeff$, will have to be taken into account. In this
section, we shall determine these corrections for arbitrary
scattering orders. A calculation for second order scattering has
been published previously in \cite{Duerr}.

The natural approach to determine $\Omeff$ is via the eigenvalues
in the Mathieu equation formalism. This approach can yield
$\Omeff$ to any desired accuracy (for constant $\Om$) by
calculating the eigenvalues of the matrix representing an
appropriately large subset of the infinite set of equations.
In this section, we are looking for an efficient iterative method
to calculate $\Omeff$, for constant as well as time-varying $\Om$.
This is, at the same time, a method for calculating the difference
of the eigenvalues $a_m-b_m$ of the Mathieu equation. This method
is based on an extension of the idea of adiabatic elimination.

We express the equation of motion Eq (\ref{momentumeqs}) as the
matrix equation
\begin{equation}
\mathcal{H}\vec{g}=i\dot{\vec{g}}
\end{equation}
where the vector $\vec g$ contains the $g_n$
\begin{equation}
\vec{g}=(\cdots,g_{-n},g_{-n+2},\cdots).
\end{equation}
In analogy to the above solution in the Mathieu equation
formalism, we are interested in a solution that is slowly varying
in time, in which the population is mainly consisting of $g_{-n}$
and $g_{+n}$. The evolution of the other states is governed by the
equation
\begin{equation}
\label{eqofmotionvector}
\tilde{\mathcal{H}}\vec{\mathcal{G}}=\vec{c}+i\dot{\vec{\mathcal{G}}},
\end{equation}
where $\vec{\mathcal{G}}$ is the vector $\vec g$ with $g_{\pm n}$
removed, $\tilde{\mathcal{H}}$ is $\mathcal{H}$ with rows and
columns of $g_{\pm n}$ removed, and
\begin{equation}\vec{c}=-\frac{1}{2}(\cdots 0,\Om g_{-n},\Om^\ast
g_{-n},0 \cdots 0,\Om g_{+n},\Om^\ast g_{+n},0 \cdots).
\end{equation}
Suppose for now that $\vec{\mathcal{G}}$ adiabatically follows
$g_{\pm n}$, i.e., $i\dot{\vec{\mathcal{G}}}\approx 0$.
$\vec{\mathcal{G}}$ thus can be expressed as functions that are
linear in $\vec{c}$, and thus $g_{\pm n}$ (but not necessarily
$\Omega$):
\begin{equation}\label{EoMG}
\vec{\mathcal{G}}=\tilde{\mathcal{H}}^{-1}\left(\vec{c}+i\dot{\vec{\mathcal{G}}}\right)\approx
\tilde{\mathcal{H}}^{-1}\vec{c}\equiv \vec{\mathcal{G}}^{(0)}.
\end{equation}
Let us define
\begin{equation}\label{Gdef}
\vec{\mathcal{G}}^{(0)}\equiv
\vec{D}^{(0)}\left(\Om,\Om^\ast\right)g_{-n}+\vec{E}^{(0)}\left(\Om,\Om^\ast\right)g_{+n},
\end{equation}
that is,
\begin{equation}\label{substitution} g^{(0)}_{-n+2m}\equiv
D^{(0)}_{m}\left(\Om,\Om^\ast\right)g_{-n}+E^{(0)}_{m}\left(\Om,\Om^\ast\right)g_{+n}.
\end{equation}
In
\begin{equation}
\label{EoMg-} i\dot{g}_{-n}\approx
\frac{\Om^\ast}{2}g^{(0)}_{-n-2}+\frac{\Om}{2}g^{(0)}_{-n+2}
\end{equation}
we apply Eq. (\ref{substitution}) to replace the $g^{(0)}_{-n\pm
2}$ and obtain the analogy to Eq. (\ref{adiabaticdiffeq}),
\begin{equation}
i\dot{g}_{-n}=\Om_{\rm ac}\left(\Om,\Om^\ast\right)g_{-n}+\frac 12
\Omeff\left(\Om,\Om^\ast\right)g_{+n},\label{g-}
\end{equation}
where
\begin{equation}
\Om_{\rm ac}=\frac{1}{2}\left(\Om^\ast D^{(0)}_{-1}+\Om
D^{(0)}_{1}\right),\quad \Omeff =\frac{1}{2}\left(\Om^\ast
E^{(0)}_{-1}+\Om E^{(0)}_{1}\right)
\end{equation}
An analogous computation leads to the expansion for
$g_{+n}$,
\begin{eqnarray}
i\dot{g}_{-n}&=&\Om_{\rm ac}g_{-n}+\frac{\Omeff}{2}g_{+n},\nonumber\\
i\dot{g}_{+n}&=&\Om_{\rm
ac}g_{+n}+\frac{\Omeff^{\ast}}{2}g_{-n}.\nonumber
\end{eqnarray}
in analogy to Eqs. (\ref{adiabaticdiffeq}), where $|\Omeff|$ is
the effective Rabi frequency. The leading order in $\Om$ of
$\Omeff$ obtained this way is identical to the one obtained in
previous section, see Eq. (\ref{Omeff}).

Although initially the fast varying $i\dot{\vec{\mathcal{G}}}$ was
set to zero by the adiabaticity assumption, we now take into
account a slowly varying part due to the adiabatic following,
which has similar time scale as the initial and final states and
thus cannot be ignored in Eq.(\ref{EoMG}). In the remainder of
this section, $i\dot{\vec{\mathcal{G}}}$ refers to this slowly
varying part only. To first order, $i\dot{\vec{\mathcal{G}}}$ can
be approximated from $\vec{\mathcal{G}}^{(0)}$:
\begin{equation}
i\dot{\vec{\mathcal{G}}}\approx
i\frac{d}{dt}\left(\vec{\mathcal{G}}^{(0)}\right).
\end{equation}
Inserting into Eq. (\ref{EoMG}),
\begin{equation}
\vec{\mathcal{G}} \approx
\vec{\mathcal{G}}^{(0)}+\tilde{\mathcal{H}}^{-1}\left(i\frac{d}{dt}
\left(\vec{\mathcal{G}}^{(0)}\right)\right).
\end{equation}
Since $\vec{\mathcal{G}}^{(0)}$ is a function of $\Om$ and $g_{\pm
n}$, the time derivatives $i\dot{g}_{\pm
n}=\frac{\Om^\ast}{2}g_{\pm n-2}+\frac{\Om}{2}g_{\pm n+2}$ contain
$\vec{\mathcal{G}}^{(0)}$ as well as still unknown corrections of
$\vec{\mathcal{G}}$. Therefore, care must be taken to properly
separate various orders of corrections. We expand
$\vec{\mathcal{G}}$ and $\dot{\vec{\mathcal{G}}}$ as
$\vec{\mathcal{G}}=\vec{\mathcal{G}}^{(0)}+\vec{\mathcal{G}}^{(1)}+\cdots$
and
$\dot{\vec{\mathcal{G}}}=\vec{\mathfrak{G}}^{(0)}+\vec{\mathfrak{G}}^{(1)}+\cdots$,
where $\vec{\mathcal{G}}^{(l)}$ ($\vec{\mathfrak{G}}^{(l)}$) are
an order of magnitude larger than $\vec{\mathcal{G}}^{(l+1)}$
($\vec{\mathfrak{G}}^{(l+1)}$) and are functions of $g_{\pm n}$
and $\Om, \dot{\Om},\ldots$:
\begin{eqnarray}\label{Dsubst}
\vec{\mathcal{G}}^{(l)}&=&\vec{D}^{(l)}(\Om, \dot{\Om},\ldots)g_{-n}\nonumber \\
&&+\vec{E}^{(l)}(\Om, \dot{\Om}, \ldots)g_{+n}.
\end{eqnarray}
We expand
\begin{eqnarray}\label{fracG}
i\dot{\vec{\mathcal{G}}}
&=&\sum^{\infty}_{l,q=0}\left(\frac{\vec{D}^{(l)}}{2}\left(\Om^\ast g^{(q)}_{-n-2}+\Om g^{(q)}_{-n+2}\right)+ig_{-n}\dot{\vec{D}}^{(l)}\right.\nonumber\\
&&\left.+\frac{\vec{E}^{(l)}}{2}\left(\Om^\ast g^{(q)}_{+n-2}+ \Om
g^{(q)}_{+n+2}\right)+ig_{+n}\dot{\vec{E}}^{(l)}\right), \nonumber
\\
i\vec{\mathfrak{G}}^{(p)}&=&\sum^{l=p}_{\substack{l=0\\q=p-l}}\left(\frac{\vec{D}^{(l)}}{2}\left(\Om^\ast g^{(q)}_{-n-2}+\Om g^{(q)}_{-n+2}\right)+ig_{-n}\dot{\vec{D}}^{(l)}\right.\nonumber\\
&&\left.+\frac{\vec{E}^{(l)}}{2}\left(\Om^\ast g^{(q)}_{+n-2}+ \Om
g^{(q)}_{+n+2}\right)+ig_{+n}\dot{\vec{E}}^{(l)}\right).
\end{eqnarray}
Eq.(\ref{eqofmotionvector}) thus becomes
\begin{equation}\label{iterative}
\tilde{\mathcal{H}}\sum^{\infty}_{p=0}\vec{\mathcal{G}}^{(p)}=
\vec{c}+i\sum^{\infty}_{p=0}\left(\dot{\vec{\mathcal{G}}}\right)^{(p)}\equiv
\vec{c}+i\sum^{\infty}_{p=0}\vec{\mathfrak{G}}^{(p)}.
\end{equation}

Since the time derivative decreases one order of magnitude for
each increase in $p$, the $(p+1)$-th order in Eq.(\ref{iterative})
is
\begin{equation}
\tilde{\mathcal{H}}\vec{\mathcal{G}}^{(p+1)}=i\vec{\mathfrak{G}}^{(p)},\
\tilde{\mathcal{H}}\vec{\mathcal{G}}^{(0)}=\vec{c}.\end{equation}
Thus,
\begin{eqnarray}\label{following}
\vec{\mathcal{G}}^{(0)}&=&\tilde{\mathcal{H}}^{-1}\vec{c}\nonumber \\
\vec{\mathcal{G}}^{(p+1)}&=&i\tilde{\mathcal{H}}^{-1}\vec{\mathfrak{G}}^{(p)}\nonumber\\
&\equiv&\vec{D}^{(p+1)}g_{-n}+\vec{E}^{(p+1)}g_{+n}.
\end{eqnarray}
We are now ready to describe an iterative procedure to obtain
$\vec{\mathcal G}$ to any desired order: We start from
$\vec{\mathcal{G}}^{(0)}$ as defined in Eq. (\ref{EoMG}). Each
component is known as a linear combination of $g_{\pm n}$, and
therefore $\vec D^{(0)}, \vec E^{(0)}$ are also known.

Now suppose that $\vec{\mathcal{G}}^{(q)}, \vec D^{(q)}, \vec
E^{(q)}$ are known for $q\leq p$. In the last Eq. (\ref{fracG}),
each $g_{-n\pm 2}^{(q)}, g_{n\pm 2}^{(q)}$ can be expressed as a
component of $\vec{\mathcal G^{(q)}}$ which, by Eq.
(\ref{Dsubst}), is known as a linear combination of $g_{\pm n}$.
We insert this into Eq. (\ref{following}) to obtain $\vec{\mathcal
G^{(p+1)}}$, again as a linear combination of $g_{\pm n}$. The
coefficients of this linear combination are the $\vec
D^{(p+1)},\vec E^{(p+1)}$. The process can now be iterated to
obtain the next higher order.

The efficiency of this method is, in part, due to the fact that
each order can be computed using the same inverse matrix
$\tilde{\mathcal{H}}^{-1}$. (Moreover, $\tilde{\mathcal H}$ is a
tridiagonal matrix, which helps in computing the inverse.)
$\Omeff$ to $M$-th order is thus obtained by plugging
$\sum^{M}_{p=0}\vec{\mathcal{G}}^{(p)}$ into Eq.(\ref{EoMg-}) and
finding the coefficient of $g_{+n}$ as in Eq.(\ref{g-}).

Note that $\mathcal{H}$ and $\tilde{\mathcal{H}}$ in principle are
infinite-dimensional. However, for obtaining the effective Rabi
frequency to order $\mathcal O\left(\Om^{2k}\right)$, it is
sufficient to include the initial and final states, states in
between, and $k$ nearest neighboring states on each side, i.e.,
$m=-k,\cdots,n+k$. Including more states yields the same result.

With this method, we explicitly calculate the effective Rabi
frequency for Bragg diffraction orders of $n\leq 29$. They can be
given as power series in $\Om$ and $\dot\Om$:
\begin{eqnarray}
\label{omeff_general} \frac{\Omeff}{\omr}&=&\frac{1}{8^{n-1}
[(n-1)!]^2}\left(\frac{\Om}{\omr}\right)^n
\left|1-\sum_{j=1}\alpha^{(2j)}_n\left(\frac{\left|\Om\right|}{\omr}\right)^{2j}\right.
\nonumber
\\ &&
\left.-\sum_{j=1}\beta^{(j)}_n\left(\frac{\dot{\Om}}{\omr\Om}\right)^{j}+\cdots\right|.
\end{eqnarray}
At first, this results in a list of numerical values for the
coefficients $\alpha$ and $\beta$ for each $n$. However, closed
expressions as function of $n$ can be found, which are listed in
appendix \ref{powerseriescoeffs}.

Eq.(\ref{omeff_general}) also allows us to give validity
conditions for the simple adiabatic elimination method presented
in Sec. \ref{adiabaticbragg}. For this to be a good approximation,
the corrections should be much less than 1. Thus, we obtain
\begin{equation}\label{validityadiabat}
\frac{n+2}{2^4(n^2-1)^2}\left(\frac{|\Om|}{\omr}\right)^2\ll 1,\
\left|\beta^{(1)}_{n}\left(\frac{\dot{\Om}}{\omr\Om}\right)\right|\ll
1.\end{equation} For large $n$, the first of these conditions
translates into $\Om\ll 4n^{3/2}\omr$, which is actually larger
than the one given by the adiabaticity condition $\om_{-n+2m} \ge
(n-1) \omr \gg \Om$.

\subsection*{Population in other states}

Summing up the population of states other than $g_{\pm n}$ at the
end of a pulse (where $g_{-n}\approx 0,g_{+n}\approx 1$), we
obtain
\begin{eqnarray}
\vec{\mathcal{G}}\cdot\vec{\mathcal{G}}^{\ast}&=&\frac{n^2+1}{2^5(n^2-1)^2}
\left(\frac{|\Om|}{\omr}\right)^2\nonumber
\\ &&+\frac{(n^4+6n^2+1)}{2^9\left(n^2-1\right)^4}
\left|\left(\frac{\dot{\Om}}{\omr^2}\right)\right|^2+\cdots.
\end{eqnarray}
The population lost into other states after the pulse is switched
off, when $\Om=\dot\Om=0$, vanishes. The method presented in this
chapter is not suitable for obtaining those losses, because the
states other than $g_{\pm n}$ have been assumed to adiabatically
follow their neighbors, and the losses are a non-adiabatic
phenomenon.

However, as long as $|g_{m\neq\pm n}|^2\ll |g_{\pm n}|^2$, the
effect of the losses on $g_{\pm n}$ and thus $\Omeff$ can be
neglected.

This iterative method, although powerful for calculating $\Omeff$,
does not approach an exact solution. It can be seen from
Eq.(\ref{g-}) that this method gives no wiggles in the sinusoidal
change of the initial or final state population, while there are
fast variations in the exact solution of a square pulse as shown
in Mathieu function section. 
However, it approaches the solution for
the initial and final state as averaged over the high-frequency
wiggles, Eq. (\ref{Mathieumean}) and thus predicts the correct
effective Rabi frequency.

\section{Adiabatic expansion}
\label{Highorderadiabat}

To investigate the losses, corrections to the adiabatic method
must be calculated. We will relabel the results $g_{\pm n}$ of the
adiabatic method as $g^{(1)}_{\pm n}$. They represent the first
order adiabatic approximation. We now want to calculate
corrections to the population of the states,
\begin{equation}
g_m=g^{(1)}_m+g^{(2)}_m+\ldots\,,
\end{equation}
where to first order only the initial and final state are nonzero.
For calculating the second order, we insert $g^{(1)}_{\pm n}$ into
Eqs. (\ref{momentumeqs}). Inserting $g^{(1)}_{\mp n}$ from Eq.
(\ref{adiabaticsolution}), we obtain population of the levels next
to the initial and final states to second order
\begin{eqnarray}
i\dot g_{-n\pm 2}&=&4(1\mp n)\om_r g_{-n\pm 2}\nonumber \\
&& +\frac12 \Om \cos\left(\frac12 \int_{-\infty}^t \Omeff(t')
dt\right) \nonumber
\\ i\dot g_{n\pm 2}&=&4(1\pm n)\om_r g_{n\pm 2}\nonumber \\ && -\frac i2
\Om \sin\left(\frac12 \int_{-\infty}^t \Omeff(t') dt\right)\,.
\end{eqnarray}
$(n>2)$. These are all states for which $g^{(2)}\neq 0$. The
process can be iterated: From the $g^{(2)}_{\pm n \pm 2}$,
corrections $g^{(3)}$ can be obtained, and so forth. Here and
throughout, we shall drop the superscript $^{(2)}$ as long as no
confusion arises.

These inhomogenous equations can be solved by standard methods,
such as variation of the constant or a Green's
function:
\begin{eqnarray} g_{-n\pm 2}(t)&=&-\frac i2
\int_{-\infty}^t dt_0\Om(t_0) \cos\left(\frac12
\int_{-\infty}^{t_0} \Omeff(t')
dt'\right)\nonumber \\ && \times e^{-4i(1\mp n)\om_r(t-t_0)} \theta(t-t_0)\,, \nonumber \\
g_{n\pm 2}(t)&=&-\frac 12 \int_{-\infty}^t dt_0\Om(t_0)
\sin\left(\frac12 \int_{-\infty}^{t_0} \Omeff(t')
dt'\right)\nonumber \\ && \times e^{-4i(1\pm
n)\om_r(t-t_0)}\theta(t-t_0) \,.
\end{eqnarray}

\subsection{Losses}

We are mainly interested in the population $g_n\equiv g_n(\infty)
$. For that, we can take out a phase factor and note that
$\theta(t-t_0)=1$. The absolute squares of these give the
population in the neighboring states. These are closed, analytic
expressions for the losses arising in the second order. As an
example for a third order correction, \begin{eqnarray}
g^{(3)}_{-n-4}(t)&=&-\frac i2\int_{-\infty}^t
dt_3\Om(t_3)e^{-8i(2+n)\omr(t-t_3)}\theta(t-t_3)\nonumber
\\ && \times \left(-\frac i2\right) \int_{-\infty}^{t_3}
dt_2\Om(t_2)e^{-4i(1+n)\omr(t_3-t_2)}\nonumber
\\ && \times \theta(t_3-t_2) \cos\left(\frac12 \int_{-\infty}^{t_2} \Omeff(t_1) dt_1\right)\,.
\end{eqnarray}

\subsection{Phase shifts}

To 3rd order, both $g_{n\pm 2}^{(2)}$ contribute to $g^{(3)}_n$:
\begin{eqnarray}\label{g3}
g^{(3)}_n& =&\frac{-1}{4}\int_{-\infty}^t
dt_3\Om(t_3)\int_{-\infty}^{t_3}dt_2\Om(t_2)\nonumber
\\ && \times[e^{-4i(1-n)\omr(t_3-t_2)}+e^{-4i(1+n)\omr(t_3-t_2)}]\nonumber \\ && \times \cos\left(\frac12
\int_{-\infty}^{t_2}\Omeff(t_1)dt_1\right)
\end{eqnarray}
[The $\theta(t_3-t_2)$ has been omitted because it is one
throughout the $t_2$ integration range.] Since this is a complex
number, the phase of $g^{(1)}_n+g^{(3)}_n$ will be shifted by some
$\Delta \phi$ relative to the adiabatic result, $g^{(1)}_n$.

An upper limit on $\Delta \phi$ can be derived as follows: We will
assume that all the losses derived in the previous section in
second order contribute to 100\% to $g_n^{(3)}$. Furthermore, we
will assume that the phase of these contributions is 90$^\circ$
shifted relative to $g_n^{(1)}$. This will result in an upper
limit on $\Delta\phi$, since in reality only a fraction of the
losses in second order will be re-introduced into the final state
to third order, and since a 90$^\circ$ phase of these
contributions is the worst case. Then,
\begin{equation}
|\Delta\phi|\leq\frac{|g^{(3)}_n|}{|g_n^{(1)}+g_n^{(3)}|}\approx
|g^{(3)}_n|\leq \sqrt{\ell}\,, \end{equation} where the total
losses are denoted $\ell$.

The equations derived in this chapter are closed expressions for
the calculation of losses and phase-shifts for realistic,
time-dependent envelope functions. We will apply them to square
and Gaussian pulses in the next two chapters.

\section{Square Pulses}

\label{squarepuls} 
\begin{figure}
\centering \epsfig{file=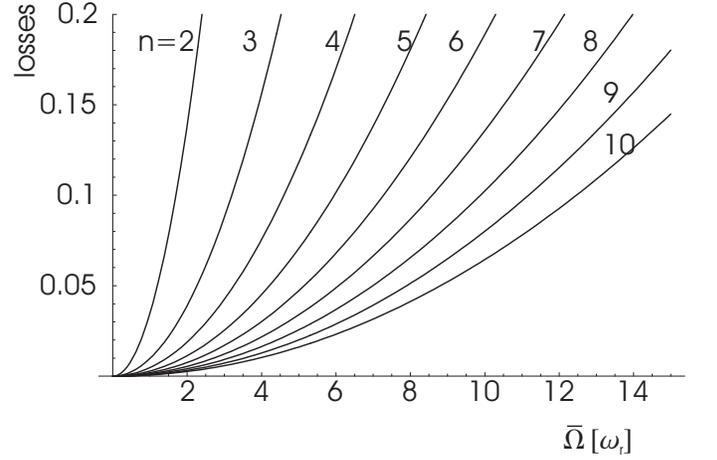,width=0.5\textwidth}
\caption{\label{lossessquare} Losses for square pulses for
$n=2,\ldots 10$ plotted versus $\bar \Om/\om_r$.}
\end{figure}

These expressions for the second order are
easily solved for square pulses having a peak two-photon Rabi
frequency of $\bar\Om$ for $0<t<T$ and zero otherwise. For the
effective Rabi frequency, we can use the result from the simple
adiabatic elimination method, because, as it will turn out, losses
become very large already in the region where this is still valid.
In this case,
\begin{eqnarray}
g_{-n\pm 2}&=&-i\frac{\bar\Om}{2}\int_0^T dt_0\cos\left(\frac{\pi
t_0}{2T}\right)e^{4i(1\mp n)\om_rt_0}\,, \nonumber \\ g_{n\pm
2}&=&-\frac{\bar\Om}{2}\int_0^T dt_0\sin\left(\frac{\pi
t_0}{2T}\right)e^{4i(1\pm n)\om_rt_0}\,.
\end{eqnarray}
The integrals can be calculated in a straightforward way; the
result can be simplified by considering the terms linear in
$\bar\Om$ only (noting that $(\omr T)^{-1}$ is proportional to
$\bar\Om^n/\omr^n$). This does not lead to reduced accuracy, since
in this second order adiabatic expansion, terms of higher order
can be assumed to be zero. We obtain
\begin{eqnarray}
g_{-n\pm 2} &=& \frac{\bar\Om}{\omr} \frac{1}{8(1\mp n)}\,, \nonumber \\
g_{n\pm 2} &=& i\frac{\bar\Om}{\omr}\frac{1}{8(1\pm n)}e^{4i(1\pm
n)\omr T}\,.
\end{eqnarray}
The losses in the final and initial state are
$\ell=|g_{-n-2}|^2+|g_{-n+2}|^2+|g_{n-2}|^2+|g_{n+2}|^2=2(|g_{n-2}|^2+|g_{n+2}|^2)$.
We obtain
\begin{equation} \ell
=\frac{1}{16}\frac{\bar\Om^2}{\om_r^2}\frac{n^2+1}{(n^2-1)^2}\,,
\end{equation}
as plotted in Fig. \ref{lossessquare}. As an example, for
$n=3,\bar\Om=3$ we have $\ell=0.088$, in excellent agreement with
the exact solution in terms of Mathieu functions (0.086).

The square-root of this is at the same time an upper limit on the
phase shift for a $\pi/2$ pulse, and indeed the exact solution as
plotted in Fig. \ref{mathieuphase} verifies this. Square pulses
are thus, unfortunately, not suitable for quasi-Bragg scattering,
because $\bar\Om$ must be reduced strongly if low losses and phase
shifts are desirable. Already the above example, with 9\% losses
and thus $\sim 0.3$\,rad maximum phase requires a $\pi$-pulse time
of about $30/\omr$, or about 3\,ms for cesium atoms. Reduction of
this to below $10^{-4}$, as required for high-precision atom
interferometers, would take a pulse time of about $10^6$ seconds.

In the next chapter, we will study pulses with a smooth envelope
function, Gaussian pulses. These will exhibit much lower losses
and phase-shifts for a given pulse duration.

\section{Gaussian pulses}
\label{Gaussianpuls}

In this section, we specialize to Gaussian pulses:
\begin{equation} \Om=\bar\Om e^{-t^2/(2\sigma^2)} \,, \end{equation} so that,
from Eq. (\ref{Omeff}), \begin{equation}
\Omeff=\bar\Om^n\left(\frac{1}{8\om_r}\right)^{n-1}\frac{1}{[(n-1)!]^2}
e^{-t^2n/(2\sigma^2)}\,. \end{equation} The case of Gaussian
pulses is more difficult. (i) One the one hand, this is because
the integrals are much harder. (ii) With Gaussian pulses, higher
Rabi frequencies can be used. Thus, we need to take into account
higher order corrections for the effective Rabi frequency.

For now specializing to $\pi$-pulses, we have the condition
\begin{equation} \label{pipulse}
\frac12 \int_{-\infty}^\infty
\Omeff(t')dt'=\frac{\bar\Om^n}{2}\left(\frac{1}{8\om_r}\right)^{n-1}\frac{\sqrt{2\pi\sigma^2}}{[(n-1)!]^2\sqrt{n}}
= \frac \pi2 \,,
\end{equation}
that we will use to determine
$\sigma$. We insert into the expressions for the amplitudes of the
neighbor states and obtain
\begin{eqnarray}\label{Gaussiang}
g_{-n\pm 2}&=&-i\frac{\bar\Omega}{2}\int_{-\infty}^\infty dt
e^{-t^2/(2\sigma^2)}
\\ &&\times \cos\left[\frac \pi
4\erp\left(\sqrt{\frac{n}{2\sigma^2}}t\right)\right]
 e^{4i(1\mp n)\om_rt}
\nonumber \\
g_{n\pm 2}&=&-\frac{\bar\Omega}{2}\int_{-\infty}^\infty dt
e^{-t^2/(2\sigma^2)}
\\ && \times \sin\left[\frac \pi 4
\erp\left(\sqrt{\frac{n}{2\sigma^2}}t\right)\right]\nonumber
 e^{4i(1\pm n)\om_rt}\,.
\end{eqnarray}
We denoted $\erp(x)=1+\Phi(x)$ and $\Phi$ is the error function
\begin{equation}
\Phi(x)\equiv \frac{2}{\sqrt{\pi}}\int_0^x
e^{-t^2}dt.
\end{equation}
For $\pi/2$ pulses, the factors of $\pi/4$ in the last three
equations will be replaced by $\pi/8$. No simpler form of this
integrals has been found. However, for $\pi$-pulses, we can use
\begin{equation}
\sin\left(\frac \pi 4+x\right) = \frac{\cos x+\sin
x}{\sqrt{2}} = \cos\left(\frac \pi 4-x\right)
\end{equation}
and $\Phi (x)=-\Phi(-x)$ to obtain $g_{-n\pm 2} = ig_{n\mp 2}^*$.

\subsubsection{Inclusion of higher-order corrections to the
effective Rabi frequency}

For Gaussian pulses, the peak effective Rabi frequency can reach a
level beyond the region of validity of Eq. (\ref{Omeff}) before
the losses as predicted from the results of the last section
become noticeable. In this region, however, the second order
adiabatic solution is still a good approximation, as long as the
losses it predicts are still low (because then, the third order
will be even lower). Therefore, we can extend the region of
validity into the interesting region, where the losses just start,
by including higher order corrections to $\Omeff$ as calculated in
section \ref{Rabifreqmethod}. In principle, of course, adiabatic
expansion to higher and higher orders is a systematic way of
obtaining results of similar and higher accuracy and
simultaneously predict the changes in the effective Rabi
frequency. However, this is very inconvenient method for Gaussian
pulses, for which the integral is hard even in the second order.

For a $\pi$ pulse we have the condition
\begin{eqnarray}
\frac12 \int_{-\infty}^\infty \Omeff(t')dt'= \frac \pi2 \,,
\end{eqnarray}
We insert Eq. (\ref{omeff_general}) and note that while the
$\alpha$ coefficients and $\beta_n^{(2)}$ are real,
$\beta_n^{(1)}$ is imaginary. $\Omeff$ is given by the absolute
magnitude of the series and we expand the absolute value as a
Taylor series in terms of the imaginary part that we truncate
after the $(\dot\Om)^2$ terms.
\begin{eqnarray}
\frac\Omeff\omr&\approx &\frac{1}{8^{n-1}
[(n-1)!]^2}\left(\frac{\Om}{\omr}\right)^n
\left[1-\sum_{j=1}\alpha^{(2j)}_n\left(\frac{\left|\Om\right|}{\omr}\right)^{2j}\right.
\nonumber
\\ &&
\left.-\tilde\beta\left(\frac{\dot\Om}{\omr\Om}\right)^2 \right],
\end{eqnarray}
where
\begin{equation}
\tilde\beta=-\frac12|\beta_n^{(1)}|^2+\beta_n^{(2)}\,.
\end{equation}
This approximation will be justified in retrospect, as we will
find that transitions having low loss satisfy the condition Eq.
(\ref{Gaussianlimit2}), which gives a lower limit on $\sigma$
\cite{remark}. At this limit, the leading neglected terms
($\beta_n^{(3)}$ and $\tilde \beta \alpha_n^{(2)}$) are starting
to become non-negligible and the comparison to a numerical
simulation shows an about 10\% error in the predicted sigma, see
Sec. \ref{accuracy}. The inclusion of more terms would reduce the
error here, but would lead to extremely lengthy expressions later.
In the region of low loss, they are negligible.

\begin{figure}
\centering \epsfig{file=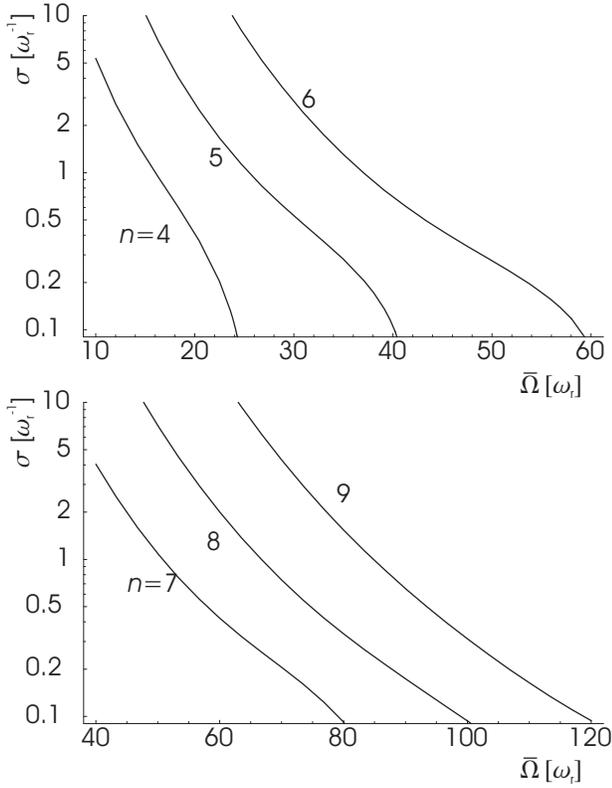,width=0.45\textwidth}
\caption{\label{gaussianresults2} $\sigma$ in units of $\omr$ for
$\pi$ pulses for Bragg diffractions of order $n=4,5,6$ (above)
$n=7,8,9$ (below) plotted versus $\bar\Om$. The effective Rabi
frequency has been inserted up to order $\bar\Om^{n+6}$.}
\end{figure}

The condition for a $\pi$-pulse reads,
\begin{eqnarray}\label{pi4ho}
\frac{\bar\Om^n}{2}
\left(\frac{1}{8\om_r}\right)^{n-1}\frac{\sqrt{2\pi\sigma^2}}{[(n-1)!]^2\sqrt{n}}
\nonumber \\
\times
\left(1-\tilde\beta\frac{1}{n\omr^2\sigma^2}-\ldots-\alpha_n^{(2)}\frac{\bar\Om^2}{\omr^2}\sqrt{\frac{n}{n+2}}
\right. \nonumber
\\ \left.
-\alpha_n^{(4)}\frac{\bar\Om^4}{\omr^4}\sqrt{\frac{n}{n+4}}-\alpha_n^{(6)}\frac{\bar\Om^6}{\omr^6}\sqrt{\frac{n}{n+6}}-
\ldots\right) =\frac\pi 2
\end{eqnarray}
From this, $\sigma$ can be computed without difficulty, see Fig.
\ref{gaussianresults2}. In the following, we take into account the
higher-order corrections up to order $\bar\Om^{n+6}$ and $1/(\omr
\sigma)^2$. For the integrals appearing in the trigonometric
functions, we find:
\begin{eqnarray}
\frac12 \int_{-\infty}^{t}\Omeff(t') dt'
=\frac{\bar\Om^n}{(8\omr)^{n-1}[(n-1)!]^2}\sqrt{\frac{\pi\sigma^2}{2n}}\nonumber
\\
\times \left[\left(1-\bar\beta\frac{1}{
n\omr^2\sigma^2}\right)\erp\left(\sqrt{\frac n 2}\frac
t\sigma\right) \right.\nonumber
\\ \left. -\sqrt{\frac{n}{n+2}}\alpha_n^{(2)}\frac{\bar\Om^2}{\omr^2}\erp\left(\sqrt{\frac{n+2}{2}}\frac
t\sigma\right)\right.\nonumber
\\ \left.-\sqrt{\frac{n}{n+4}}\alpha_n^{(4)}\frac{\bar\Om^4}{\omr^4}\erp\left(\sqrt{\frac{n+4}{2}}\frac
t\sigma\right)\right. \nonumber \\ \left.
-\frac{\bar\beta}{\omr^2\sigma^3}\sqrt{\frac{2}{n\pi}}te^{-\frac{nt^2}{2\sigma^2}}\right]\,.
\end{eqnarray}
Making the same assumptions as above, we can neglect the
$1/\sigma^3$ term. Using Eq. (\ref{pi4ho}), we can write
\begin{widetext}
\begin{eqnarray*}\label{long}
\sin\left[\frac12 \int_{-\infty}^{t}\Omeff(t') dt'\right]\nonumber \\
=\sin\left[\frac \pi 4\frac{\left(1-\bar\beta\frac{1}{
2n\omr^2\sigma^2}\right)\erp\left(\sqrt{\frac n 2}\frac
t\sigma\right)-\sqrt{\frac{n}{n+2}}\alpha_n^{(2)}\frac{\bar\Om^2}{\omr^2}\erp\left(\sqrt{\frac{n+2}{2}}\frac
t\sigma\right)-\sqrt{\frac{n}{n+4}}\alpha_n^{(4)}\frac{\bar\Om^4}{\omr^4}\erp\left(\sqrt{\frac{n+4}{2}}\frac
t\sigma\right)}{
1-\bar\beta\frac{1}{n\omr^2\sigma^2}-\alpha_n^{(2)}\frac{\bar\Om^2}{\omr^2}\sqrt{\frac{n}{n+2}}
-\alpha_n^{(4)}\frac{\bar\Om^4}{\omr^4}\sqrt{\frac{n}{n+4}}}\right]
\end{eqnarray*}
\end{widetext}
The functions $\erp(\sqrt{n/2}t), \erp(\sqrt{(n+2)/2}t), \ldots$,
although similar, show an increasing slope with respect to $t$. As
they are integrated over, and as $\sqrt{(n+2)/2}\approx
\sqrt{n/2}$, this is insignificant for large $n$ and we can take
these functions to be equal. We then recover Eq.
(\ref{Gaussiang}), the only difference being the replacement of
$\sigma$ by the solution of Eq. (\ref{pi4ho}). It remains to
actually calculate the integral.

This integral is a function of $n$ and $\sigma$. $\sigma$, in
turn, is determined by $\bar\Om$ for $\pi$ or $\pi/2$ pulses.
Unfortunately, the integral cannot be solved exactly. The general
structure of the integrand is a periodic oscillating function
$\exp[-i\omr(1\pm n)t]$ times an envelope that is peaked near
$t=0$. For $g_{n\pm 2}$, the peak is limited on the right mainly
by the Gaussian $\exp[-t^2/(2\sigma^2)]$ and on the left by the
sine of the error function. For low Rabi frequencies, $\sigma$ is
high and thus the peak is broad compared to the period $1/[(n\pm
1)\omr]$ of the oscillating function, which thus averages out to a
very low value. When the peak width becomes comparable to the
period, however, this is no longer the case and the value of the
integral will increase. Since we are interested in the region
where the losses are nonzero, but still low, it is sufficient to
solve the integral with a method that is valid in this region. The
method of steepest descent, or ``saddle-point" method is suitable.
As derived in Appendix \ref{saddlepointsection}, neglecting a
phase factor that is of no consequence unless one wants to proceed
to higher orders,
\begin{equation}
g_{n\pm 2}=-\bar\Om D \sqrt{2\sigma^2} \exp\left\{-8(1\pm
n)^2\frac{\sigma^2\om_r^2}{n^{1/3}\Gamma}\right\} =ig_{-n\mp
2}^*\,,
\end{equation}
see Eq. (\ref{gaussianneighbors}). Here, $\Gamma\approx1.64874$
and $D$ is a factor of the order one that is given in Eq.
(\ref{Bdef}) and plotted in Fig. \ref{Vorfaktoren}.

\begin{figure}
\centering \epsfig{file=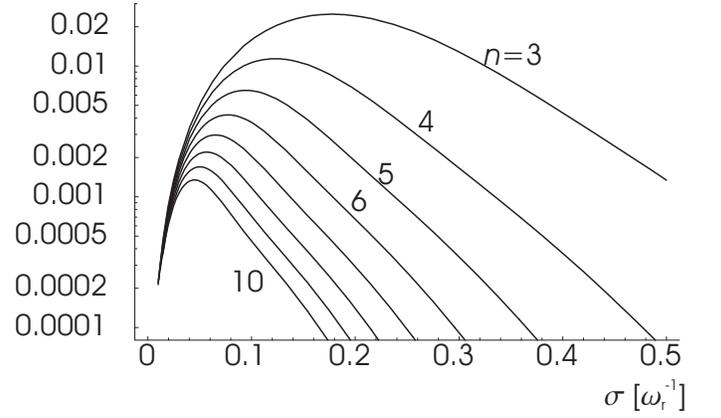,width=0.5\textwidth}
\caption{\label{numerical} Numerical calculation of the integral
$\int_{-\infty}^\infty
\exp[-t^2/(2\sigma^2)-i\omr(1-n)t]\erp\{(\pi/4)\sin[\sqrt{n/2}(t/\sigma)]\}dt$
as plotted versus $\sigma$. Parameter: $n=3,4,\ldots 10$. $\sigma$
is in units of inverse recoil frequencies.}
\end{figure}

Alternatively, the integral can be computed numerically, see Fig.
\ref{numerical}. Accurate numerical computation, however, becomes
difficult when the losses are low, because then the integrand will
make very many oscillations. The saddle point method, however,
works well especially in this region. Comparison of both shows
good agreement for the range we are interested in, where the
losses are lower than about 10\%.

As shown in Appendix \ref{saddlepointsection}, the results can be
easily adapted to $\pi/2$ pulses if $\Gamma$ is replaced by
$\Gamma_{\pi/2}=1.5043$ and $D$ by $D_{\pi/2}$, which is also
plotted in Fig. \ref{Vorfaktoren}. Thus, the losses for $\pi$ and
$\pi/2$ pulses are essentially the same.

\subsection{Losses}\label{gaussianlosses}

\begin{figure}
\centering \epsfig{file=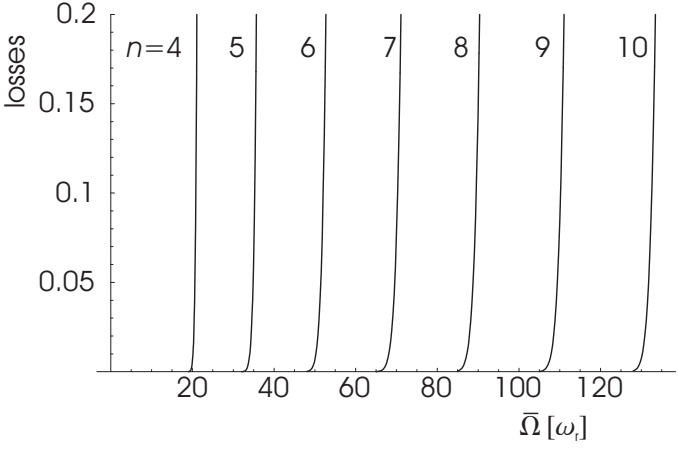,width=0.5\textwidth}
\caption{\label{gaussianresults} Losses for Gaussian $\pi$ pulses
for Bragg diffractions of order $n=4$ to 10 plotted versus
$\bar\Om$. The effective Rabi frequency has been inserted up to
order $\bar\Om^{n+6}$.}
\end{figure}

The losses in the final and initial state are
$\ell=|g_{-n-2}|^2+|g_{-n+2}|^2+|g_{n-2}|^2+|g_{n+2}|^2=2(|g_{n-2}|^2+|g_{n+2}|^2)$.
Since the terms with the plus sign in the exponential can be
neglected, we obtain a simple form
\begin{equation}
\label{Gaussianlosses} \ell = 4(\bar\Om D\sigma)^2
e^{-16(1-n)^2\sigma^2\om_r^2/(n^{1/3}\Gamma)}\,.
\end{equation}
For losses below the $10^{-2}$ level, we have the condition (the
factors outside the exponential are of order one)
\begin{equation}
\label{Gaussianlimit2}
\sigma\om_r>\frac{\sqrt{2\Gamma\ln(2)}n^{1/6}}{4(n-1)}\approx
0.38\frac{n^{1/6}}{n-1}\,.
\end{equation}
Since the losses decrease rapidly with longer $\sigma$, very low
losses can be reached: For example, if the theoretical losses
should be lower than $10^{-10}$ (which is then clearly negligible
compared to losses due to technical influences),
\begin{equation}
\label{Gaussianlimit} \sigma\om_r>  1.5\frac{n^{1/6}}{n-1}\,.
\end{equation}
We can now compute the losses and pulse durations as functions of
$\bar\Om$, see Figs. \ref{gaussianresults} and
\ref{gaussianresults2}.

\subsection{Phase shifts}\label{gaussianphase}

As discussed, the parasitic phase shifts in radians are at most
equal to the square-root of the losses, as computed in the
previous section. Since the losses are such a steep function of
$\bar\Om$, slightly reducing $\bar\Om$ below the level where the
losses become appreciable will essentially reduce the phase shifts
to negligible levels. For example, if Eq. (\ref{Gaussianlimit}) is
satisfied, these theoretical shifts will be below $10^{-5}$.

For Gaussian pulses with an appropriate choice of $\bar\Om$, the
theoretical losses considered here are so low that they are
negligible in practice. The practical losses will then be
dominated by issues such as single-photon excitation, finite laser
beam size, finite size and temperature of the atomic cloud,
wavefront distortions, and other things. These losses are based on
entirely different mechanisms and will therefore not necessarily
lead to a phase shift as the theoretical losses calculated in this
paper.

\subsection{Truncated Gaussians}

Any experimental realization of the Gaussian must be truncated
somewhere. Here, we consider the modifications of the above
considerations that arise if the Gaussian is truncated on the
right side (only) at $t=\tau\gg \sigma$. The integrations in Eq.
(\ref{Gaussiang}) will now run from $-\infty$ to $\tau$. It is
convenient to write the final state after a truncated Gaussian
pulse as $g_{-n\pm 2,n\pm2}+g^\tau_{-n\pm2,n\pm2}$, where
$g_{-n\pm 2,n\pm2}$ is given by Eq. (\ref{Gaussiang}) and
\begin{eqnarray}
g_{-n\pm 2}^\tau&=&i\frac{\bar\Omega}{2}\int_{\tau}^\infty dt
e^{-t^2/(2\sigma^2)}
\\ &&\times \cos\left\{\frac \pi 4
\erp\left(\sqrt{\frac{n}{2\sigma^2}}t\right)\right\}
 e^{4i(1\mp n)\om_rt}
\nonumber \\
g_{n\pm 2}^\tau&=&\frac{\bar\Omega}{2}\int_{\tau}^{\infty} dt
e^{-t^2/(2\sigma^2)}
\\ && \times \sin\left\{\frac \pi 4
\erp\left(\sqrt{\frac{n}{2\sigma^2}}t\right)\right\}\nonumber
 e^{4i(1\pm n)\om_rt}\,.
\end{eqnarray}
Since $\tau\gg \sigma$, we can use $\Phi(t)\approx
1-e^{-t^2}/(\sqrt{\pi} t)$ and approximate the trigonometric
functions to leading order of the argument. We obtain
\begin{eqnarray}
g_{-n\pm
2}^\tau&=&i\frac{\bar\Omega}{8}\sqrt{\frac{2\pi\sigma^2}{n}}\int_{\tau}^\infty
\frac{dt}{t}e^{-(n+1)t^2/(2\sigma^2)+4i(1\mp n)\om_rt}\nonumber
\\
g_{n\pm 2}^\tau&=&-\frac{\bar\Omega}{2}\int_{\tau}^{\infty} dt
e^{-t^2/(2\sigma^2)} e^{4i(1\pm n)\om_rt}\,.
\end{eqnarray}
For $t\geq\tau\gg\sigma$, the Gaussian $e^{-t^2/(2\sigma^2)}$ (and
{\em a fortiori} $e^{-(n+1)t^2/(2\sigma^2)}$) is a very steep
function of $t$, which is non-negligible only in a small region
above $\tau$. In fact, we can assume that the other functions in
the integral are constant and have the values they take at
$t=\tau$. The integrals then reduce to error functions, which in
turn can be replaced by the asymptotic expression for large
arguments. We obtain
\begin{eqnarray}
g_{-n\pm
2}^\tau&=&-i\frac{\bar\Om}{4}\frac{\sigma^3}{\sqrt{n}(n+1)\tau^2}e^{-(n+1)\tau^2/(2\sigma^2)}e^{4i(1\mp
n)\omr\tau}\nonumber
\\
g_{n\pm
2}^\tau&=&\frac{\bar\Om\sigma^2}{\sqrt{\pi}\tau}e^{-\tau^2/(2\sigma^2)}e^{4i(1\pm
n)\omr\tau}\,.
\end{eqnarray}
Truncation on the left interchanges the expressions for
$g^\tau_{n\pm2}$ and $g^\tau_{-n\pm2}$.

For an example, consider $\tau=\sigma \sqrt{-2\ln(\eta)}$ is
chosen such that the truncation happens at a fraction $\eta$ of
the peak amplitude. Obviously, the smaller $\eta$, the smaller the
effect of truncation, which also decreases for large $n$, (as
expected, as the Gaussian to the power of $n$ has a narrower
peak). Consider, for example, $\eta=1/100$ (i.e., $\tau\approx
3\sigma$) and $n=5$. Then, $|g_{-n\pm 2}^\tau|=3.9\times
10^{-15}\bar\Om\sigma$. That means, the effects of truncation can
be reduced to negligible magnitude.

In order to keep phase-shifts low, it is desirable to choose
$\tau$ such that $4\omr\tau=2\pi,4\pi,\ldots$, in which case the
amplitudes for the neighbor states are real, i.e., there is no
extra phase shift due to the truncation. On the other hand,
purposely truncating the pulses at a variable $\tau$ thus provides
a means of testing the influence of these phase shifts in
experiments.

\subsection{Accuracy of the estimates and higher-order effects}
\label{accuracy}

The losses are very steep functions of $\bar\Om$. Therefore, the
error in the predicted loss for a given Rabi frequency is much
larger than the error in the inverse function, the Rabi frequency
for a given loss. To confirm these results, we numerically
integrate the Schr\"{o}dinger equation in the momentum
representation. We include sufficiently many ``outer" momentum
states until the results are essentially unaffected by including
more (usually, 5 are sufficient). The simulation program itself is
checked for constant Rabi frequency against the exact solution in
terms of Mathieu functions.

\begin{figure}
\centering \epsfig{file=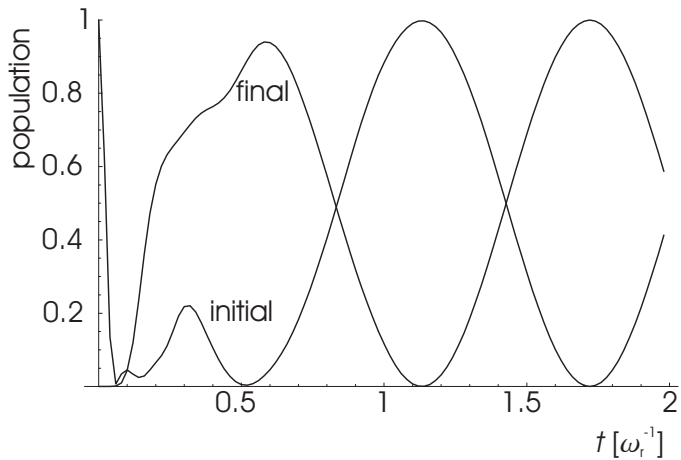,width=0.5\textwidth}
\caption{\label{simul} Simulated populations of the initial and
final state for $n=5, \bar\Om=38$ plotted versus $\sigma$ in units
of inverse recoil frequencies.}
\end{figure}

The result for Gaussian pulses with $n=5$ and $\bar\Om=38\omr$ is
shown in Fig. \ref{simul}. For $\sigma\lesssim 0.1/\omr$, most of
the population is still left in the initial state,
$|g_{-5}|^2\approx 1$. For slightly larger $\sigma$, the
population gets driven out of the initial state. However, the
transfer efficiency to the final state $g_5$ is very low, as
expected in the Raman-Nath regime. When the pulse gets longer, the
system enters the quasi-Bragg regime and the losses become quite
low. At $\sigma=0.52/\omr$, the population of the initial state
has a minimum of $|g_{-5}|^2\approx 0.0035$ whereas  the
population of the final state $|g_5|^2\approx 0.89$. The remaining
11\% of the population are lost into other states. At
$\sigma\approx 0.58/\omr$, the final state has a first maximum of
$|g_5|^2\approx 0.94$; at this time, $|g_{-5}|^2\approx 0.029$,
i.e., about 6\% are lost into other orders. Thus, in this regime,
the minimum of the initial and the maximum of the final state do
not coincide. For even larger $\sigma$, the losses become
negligible and the system performs Pendell\"osung oscillations
just as in the Bragg regime, although the adiabaticity criterion
is still violated.

For comparison, the second order adiabatic theory predicts about
6\% losses for $\bar\Om=35$, rather than 38. The bulk of the 10\%
discrepancy is due to the remaining error in the calculation of
$\sigma$ from Eq. (\ref{pi4ho}). Another prediction of the theory
that is confirmed by the simulation is the symmetry of the losses
$g_{n\pm 2}=-g_{-n\mp 2}^*$.

To third order, a part of the population that is lost to the
neighbor states to second order returns to the initial and final
state in the third order. This has two main effects: First, it
reduces the predicted losses. Second, since the returning
population will have a different time dependence for the initial
and final state, it can account for the difference of the $\sigma$
that gives maximum population in the final state versus the one
that gives a minimum in the initial state. We will not consider
this.

\subsection{Practical example}

Suppose we want to achieve $2n=16$-photon Bragg diffraction with
the lowest possible time compatible with losses below 1\%. Eq.
(\ref{Gaussianlimit2}) gives a minimum $\sigma\gtrsim
0.08\omr^{-1} \approx 6\,\mu$s. This is substantially faster than
allowed by the adiabaticity limit Eq. (\ref{adiabaticityBragg}),
which would require $\Omeff \ll 0.007\omr\approx 2\pi \times
14\,$Hz, i.e., transition times on the order of 0.1\,s. For
applications in precision atom interferometry, where the unwanted
phase shift, that is estimated as the square-root of the losses,
must be low, we might want to have theoretical losses as low as
$10^{-10}$ (even if then additional losses due to technical
reasons will be much larger). This can be obtained with $\sigma
\gtrsim 0.3\omr^{-1}\approx 24\,\mu$s, according to Eq.
(\ref{Gaussianlimit}). For this case, we can read off a required
peak two-photon Rabi frequencies of $\bar\Omega\sim 80\omr$ from
Fig. \ref{gaussianresults2}.

For an explicit example, consider cesium atoms driven on the
$6^2S_{1/2}, F=4,m_F=0 \rightarrow 6^2P_{3/2}$ (D2) transition
which has a wavelength of 852\,nm, with a detuning of 10\,GHz. The
relevant data of that transition is found in Ref. \cite{Steck}. To
determine the two-photon Rabi frequency, we neglect the hyperfine
splitting of the excited state (which is on the order of a few
100\,MHz) and sum over the excited states $|F'=5,m_F=1\rangle$,
$|F'=4,m_F=1\rangle$, and $|F'=3,m_F=1\rangle$. From summing the
matrix elements, we obtain a two-photon Rabi frequency of
$\bar\Omega=\frac 23 (I/I_{\rm sat})\frac 14(\Gamma^2/\delta)$, or
$\bar\Omega/\omr=0.20 I/($mW/cm$^2$), where $I_{\rm
sat}=1.1\,$mW/cm$^2$.

If we have, for example, laser beams with a Gaussian waist of
$w_0=0.6\,$cm, the intensity at the center is $I=2P/(\pi w_0^2)$,
where $P$ is the total power. Inserting, we obtain
$\bar\Omega/\omr=0.35 P/$mW. Thus, a power of about 250\,mW is
necessary to reach $\bar\Omega/\omr\simeq 88$.

\section{Discussion, Summary, and Outlook}

We have given an analytic theory of quasi-Bragg scattering. This
is the range of short interaction times for which the usual
adiabatic theory of Bragg scattering breaks down. Thus, population
shows up in scattering orders other than those allowed by the
Bragg condition. However, we find that this population can still
be extremely low provided that the interaction is switched on and
off with a smooth envelope function.

For calculating these effects, we introduced a new method for
solving the Schr\"odinger equation, adiabatic expansion. The
Schr\"odinger equation in momentum space is a coupled system of
differential equations describing the population in the different
momentum states. A first approximate solution is obtained in the
usual way by adiabatically eliminating the states between the
initial and the final state. We then re-insert this first-order
solution into the Schr\"odinger equation. This results in
inhomogenous differential equations for the momentum states next
to the initial and final state. Thus, we obtain a second-order
solution. While this process can be iterated, the second order is
sufficient for our purposes. Unlike previous approaches, this
method allows us to treat arbitrary diffraction orders and
envelope functions. We expect that this method is useful for
obtaining higher accuracy results in all situations that are
usually treated by adiabatic elimination.

We also present an efficient method for calculating the effective
Rabi frequency, which is related to the eigenvalues of Mathieu
functions. Closed expressions are obtained for the Rabi frequency
up to eighth-order corrections for arbitrary Bragg diffraction
order.

We treat diffraction with a scatterer having square and Gaussian
envelope functions as examples. Square envelopes lead to a high
loss of the population into undesired momentum states (as
experimentally observed, e.g., by \cite{Keller}). Moreover, due to
couplings of the desired output state with the loss channels, the
output state becomes phase-shifted as a function of the
interaction strength and -duration. On the other hand, Gaussian
envelopes can lead to extremely low losses, and hence phase
shifts, even for pulse times that substantially violate the
adiabaticity criterion. The effects of truncation of the Gaussian
to a finite waveform can be made negligible by suitably chosen
truncation. Comparison to a numerical integration of the
Schr\"odinger equation verifies that the second order is
sufficient for our purposes.

While these results are important in all situations were Bragg
diffraction is applied (like acousto-optic modulators), we focus
on the context of atom interferometry. For example, some present
high-precision atom interferometers demand phase errors
$\Delta\phi$ below $10^{-5}$\,rad \cite{Paris}. For Gaussian
pulses, the total losses $\ell$ are given by Eq.
(\ref{Gaussianlosses}) and $\Delta\phi\sim \sqrt{\ell} \leq
10^{-5}$ can be satisfied by operating with a $\sigma$ longer than
about $1.5/\omr$, see Eq. (\ref{Gaussianlimit}). The strong
dependence of $\ell$ on $\bar\Om$ means that a slight reduction of
$\bar\Om$ corresponds to a strong reduction of $\Delta\phi$. On
the other hand, for square pulses, $\sqrt{\ell}\leq 10^{-5}$ can
only be fulfilled for $\bar\Om\ll\omr$, which will in practice
mean unrealistically long transition times.

While atom interferometry is a field to which our calculations
are, we hope, useful, in this work we restricted attention to one
single beam splitter. The application of the results to a full
interferometer, e.g., in the Ramsey-Bord\'{e} or other geometries,
is a matter that we did not consider. While we considered the
phase shift arising within one beam splitter, the signal of an
atom interferometer is given by the total amplitude of all
interfering momentum states. In certain situations, the
neighboring momentum states (considered as ``losses" in here)
contribute to this, leading to an apparent distortion of the
interference fringes. This has been considered for the Raman-Nath
regime in \cite{Dubetsky2,Dubetsky3}. An upper limit for this can
be given by $\sqrt{\ell}$, by assuming that all the lost
population interferes and conspires to produce the maximum effect.
Thus, operation in the quasi-Bragg regime is suitable for reducing
this contribution. Moreover, an actual atom interferometer
consists of more than one beam splitter. Since the undesired
output states of the first one have different momentum, they will
not be addressed by the subsequent (velocity-selective) beam
splitters. This strongly reduces this effect. A further reduction
is possible by choosing the geometry of the interferometer such
that the undesired output states do not interfere.

\acknowledgments We would like to thank an anonymous referee, as
well as B. Dubetsky, T.W. H\"ansch, and Q. Long for discussions.
H.M. thanks the Alexander von Humboldt-foundation for support in
the initial phase of this work. This material is based upon work
supported by the National Science Foundation under Grant No.
0400866, the Air Force Office of Scientific Research, and the
Multi-University Research Initiative.

\appendix

\section{Coefficients of a power series expansion of the Mathieu eigenvalues}
\label{powerseriescoeffs}

We start from the numerical values obtained in Sec.
\ref{Rabifreqmethod} and make a polynomial ansatz
$D_n=(n+k_1)(n+k_2)\ldots$ for the denominator of the
$\alpha^{(2j)}_n$. The coefficients $k_i$ are then obtained by
factoring the denominators into primes. If, for example, the prime
$7$ first appears at $n=7-1$ and $n=7+1$, we choose one of the
$k_i$ equal to $-1$ and one to $+1$. This leads to a guess for the
denominator. Subsequently, an ansatz is made for the numerators
$N_n$ by a polynomial of $K$th order. The coefficients are
determined by solving $\sum^{K}_{k=0} a_k n^k=N_n$ for the
coefficients $a_k$. The resulting expression is then confirmed by
direct comparison to the above calculation. As it turns out, the
$\alpha_n^{(2j)}$, $n=1,2,\ldots N$ can be given by polynomials of
relatively low order compared to $N$. This suggests that the
polynomials may turn out to be exact expressions for all $n$,
although we have made no attempt to proof this conjecture.

The functional forms of $\beta^{(1)}_{n}$ and $\beta^{(2)}_{n}$
are obtained as follows. For $\beta^{(1)}_{n}$, a factorial
denominator $(n-1)!$ is found, and then a recursive relation of
the numerators is observed. After simplifying the relation, the
functional form is obtained and tested with larger $n$'s. For
$\beta^{(2)}_{n}$, $\om_{-n+m}$ in $\tilde{\mathcal{H}}$ are kept
not evaluated, and a pattern of the appearance of the $\om$'s in
$\beta^{(2)}_{n}$ is found. The pattern is then simplified with
$\om$'s evaluated as $\om_{-n+2m}=m(n-m)\omr$ thus yields the
form.

As a result, we obtain the following closed expressions for the
coefficients:
\begin{eqnarray}
\alpha^{(2)}_{n}=\frac{n+2}{2^4(n^2-1)^2},\quad
\alpha^{(4)}_{2}=\frac{-11141}{7077888}, \nonumber \\
\alpha^{(4)}_{n>2}=-(n+4)(4n^5-15n^4-32n^3+12n^2\nonumber
\\+64n+111)/[2^{11}(n^2-1)^4(n^2-4)^2],\nonumber
\end{eqnarray}
\begin{equation}
\alpha^{(6)}_{2}=\frac{1086647}{9555148800}, \quad
\alpha^{(6)}_{3}=\frac{-872713}{1087163596800},\nonumber
\end{equation}
\begin{eqnarray}
\alpha^{(6)}_{n>3}=(n + 6)(4n^{10} - 45n^9 + 76n^8 + 846n^7
\nonumber \\ + 484n^6 - 3960n^5 - 14824n^4 - 3078n^3  \nonumber
\\+ 31904n^2 + 26973n +30740)\nonumber \\ /[3!\ 2^{14}(n^2 -
1)^6(n^2 - 4)^2(n^2 -9)^2], \nonumber
\end{eqnarray}
\begin{eqnarray}
\alpha^{(8)}_2&=&\frac{-20778032863}{2254342434324480},\nonumber \\
\alpha^{(8)}_3&=&\frac{16738435813}{2727476031651840000}, \nonumber \\
\alpha^{(8)}_4&=&\frac{-218963004049}{2301307901706240000000},\nonumber
\end{eqnarray}
\begin{eqnarray}
\alpha^{(8)}_{n>4}=-(n + 8)(5191881936 +5562082816 n \nonumber\\
+9349559664 n^2 + 953069376 n^3 - 6361852029 n^4 \nonumber \\-
3794628570 n^5 - 375347622 n^6 + 1490778180 n^7\nonumber\\
+888115497 n^8-172767750 n^9  - 198150468 n^{10} \nonumber \\ -
10966056 n^{11} +16309509 n^{12} + 3762906 n^{13} \nonumber \\-
472854 n^{14} - 257532 n^{15} +
11487 n^{16}+ 4998 n^{17} \nonumber \\ - 720 n^{18} + 32 n^{19}) \nonumber \\
/[4! 2^{21} (n^2 - 1)^8 (n^2 - 4)^4 (n^2 - 9)^2 (n^2 - 16)^2],
\end{eqnarray}
\begin{eqnarray}
\beta^{(1)}_{n}&=&\frac{i}{4}\sum_{k=1}^{n-1}\frac{1}{k}\equiv
\frac{i}{4}H_{n-1}, \nonumber \\
\beta^{(2)}_{n}&=&\frac{1}{16}\left[\left(1-\frac{2}{n}\right)
H_{n-1}^2-\sum_{k=2}^{n-2}\frac{n-k-1}{k(n-k)}H_{k-1}\right].
\end{eqnarray}
At present, we have proved these expressions for $n<30$ by direct
computation, as explained above. The expressions have also been
checked against the differences $a_n-b_n$ of the eigenvalues of
Mathieu equation, as far as they are available from
\cite{Abramowitz}.

\section{Mean of the eigenvalues $a_n, b_n$ of Mathieu
functions}

$\Om_{\rm ac}$, that was introduced in Eq. (\ref{g-}), is the
common energy shift of the effective two-level system, which thus
corresponds to the mean $(a_n+b_n)/2$ of characteristic values for
real and constant $\Om$. Thus, by taking into account of the
difference in energy reference in Eq.(\ref{Mathieu}) and
$\mathcal{H}$, the mean of $a_n,b_n$ can be obtained. As a result,
both $a_n$ and $b_n$ are known independently.

In fact, for real constant $\Om$, every term in the power series
expansion for the mean value as a function of $n$ can be obtained
efficiently. The result is valid for all $n$ as long as the
denominator of the coefficients listed below do not vanish. This
method is used to calculate the power series expansion up to the
14th order in the parameter $q$ as introduced in Eq. (\ref{defq})
on a personal computer in 10 minutes, and the result agrees with
Abramowitz and Stegun \cite{Abramowitz}, who list the terms up to
order $q^6$. The subsequent terms are as follows: The coefficient
of $q^8$ is
\begin{eqnarray*}
(274748+827565n^2+64228n^4-140354n^6+9144n^8\nonumber
\\ +1469n^{10})/[2^{21}(n^2 -1)^7(n^2-4)^3(n^2-9)(n^2-16)];
\end{eqnarray*}
of $q^{10}$,
\begin{eqnarray*}
(4453452+20651309n^2+13541915n^4-2844430n^6\nonumber
\\ -1039598n^8+69361
n^{10}+4471n^{12})\nonumber
\\ /[2^{24}(n^2-1)^9(n^2-4)^3(n^2-9)(n^2-16)(n^2-25)],
\end{eqnarray*}
of $q^{12}$,
\begin{eqnarray*}
(1155192131376+6474038960008n^2\nonumber \\ +4470328527807 n^4-
3667584923421 n^6 \nonumber \\- 518221243968 n^8 +
604333473552n^{10}\nonumber
\\ - 86398056330 n^{12} - 5960020482 n^{14} \nonumber \\+
2031809256 n^{16} - 119406048 n^{18} -131341 n^{20} \nonumber \\+
121191
n^{22})/[2^{30}(n^2 - 1)^{11}(n^2 - 4)^5(n^2 - 9)^3\nonumber \\
\times (n^2-16)(n^2-25)(n^2-36)],
\end{eqnarray*}
and of $q^{14}$,
\begin{eqnarray*}
(41218724372688 + 302286294466120 n^2 \nonumber \\ +
433922366490105 n^4 - 32881683767026 n^6\nonumber \\  -
136265027585703 n^8 + 21249098173752 n^{10}\nonumber
\\ +7968113847666 n^{12} -
1857745713708 n^{14}\nonumber \\ + 30568658442 n^{16} +15106984464
n^{18} \nonumber \\ -
880061323 n^{20} + 3059598 n^{22} + 441325 n^{24})\nonumber \\
/[2^{33}(n^2 - 1)^{13}(n^2 - 4)^5(n^2 - 9)^3\nonumber \\ \times
(n^2 - 16)(n^2-25)(n^2-36)(n^2-49)].
\end{eqnarray*}
This is the first time the close form higher order power series
expansion terms of the mean characteristic values are reported.

\section{Calculation of the integral}\label{saddlepointsection}

\begin{figure}
\epsfig{file=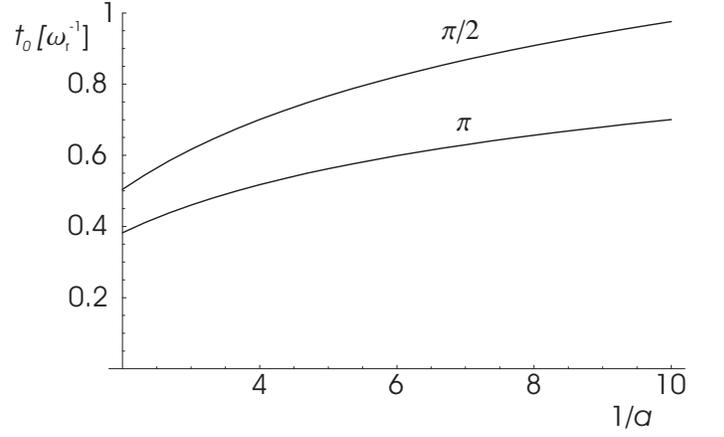,width=0.5\textwidth}\caption{\label{tNull}
$t_0$ for $\pi$ (lower graph) as well as $\pi/2$ (upper graph)
pulses, plotted versus $1/a$.}
\end{figure}

In this appendix, we calculate the integral
\begin{equation} \int_{-\infty}^\infty dt e^{-
t^2/(2\sigma^2)}\sin\left[\frac \pi 4
\erp\left(\nu\sqrt{\frac{n}{2}}\frac t\sigma
\right)\right]\nonumber
 e^{4i(1\pm n)\om_rt}\,. \end{equation}
We substitute $\nu\sqrt{n/2} (t/\sigma)$, to bring the integral
into the form
\begin{eqnarray}
g_{n\pm 2}&=&-i\sqrt{\frac{2\sigma^2}{n}}\frac{\bar\Omega}{2\nu}
g_\pm
\end{eqnarray}
where
\begin{equation} g_\pm \equiv \int_{-\infty}^\infty e^{-a
t^2-ibt}\sin\left(\frac\pi 4\erp(t)\right)dt
\end{equation}
and $a=1/n$, $b=-4(1\pm n)\sqrt{2\sigma^2/n}\om_r/\nu$. A useful
approximation is the saddle point method. We write
\begin{eqnarray} g_\pm &=&\int_{-\infty}^\infty e^{f(t)-ibt} dt \nonumber \\
f&=&-at^2+\ln\left(\sin\left[\frac\pi 4\erp(t)\right]\right)\,.
\end{eqnarray}
We expand into a Taylor Series near the maximum of $f$ at $t_0$,
\begin{equation} g_\pm=\int_{-\infty}^\infty \exp\left\{f(t_0)+\frac 12
f''(t_0)(t-t_0)^2+\ldots-ibt\right\} dt\,.
\end{equation}
Thus,
\begin{equation}
g=\sqrt{\frac{2\pi}{f''(t_0)}}e^{f(t_0)-ibt_0+\frac{b^2}{2f''(t_0)}}
\,.
\end{equation}
For finding $t_0$, we calculate \begin{equation} \label{t0}
f'(t_0)=-2at_0+\frac{\sqrt{\pi}}{2}e^{-t_0^2}\cot\left[\frac\pi
4\erp(t_0)\right]=0 \,. \end{equation} $t_0$ can be calculated
numerically and is plotted versus $1/a$ in Fig. \ref{tNull}.

\begin{figure}
\centering \epsfig{file=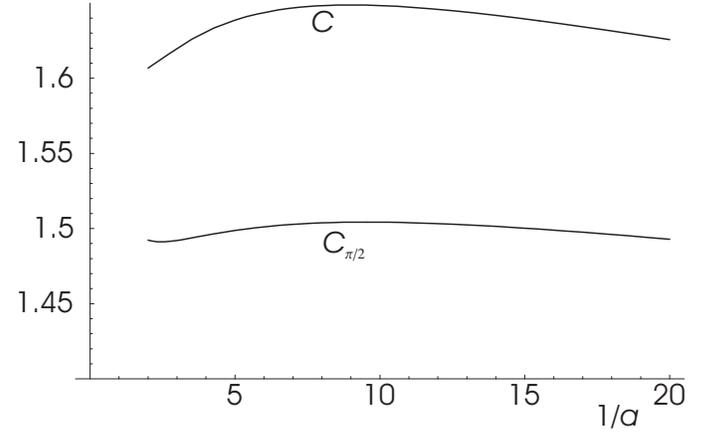,width=0.5\textwidth}
\caption{\label{C} $C$ (upper graph) and $C_{\pi/2}$ (lower graph)
plotted versus $1/a$.}
\end{figure}

For the second derivative, we calculate
\begin{eqnarray}
f''(t)&=&-2a-
e^{-t^2}\sqrt{\pi}t\cot\left[\frac\pi 4\erp(t_0)\right]\nonumber \\
&& -\frac\pi4 e^{-2t^2}\csc^2\left[\frac\pi 4\erp(t_0)\right] \,.
\end{eqnarray}
To obtain a simplified expression for $f''(t_0)$, we derive from
Eq. (\ref{t0}):
\begin{eqnarray}
e^{-t_0^2}\cot\left[\frac\pi
4\erp(t_0)\right]&=&\frac{4at_0}{\sqrt{\pi}} \nonumber \\
\csc^2\left[\frac\pi4
\erp(t_0)\right]&=&1+\frac{16}{\pi}a^2t_0^2e^{2t_0^2}\,.
\end{eqnarray}
For obtaining the latter result, $\csc^2x=1+\cot^2x$ has been
used. We obtain $ f''(t_0)=-2Ca^{2/3}$ where \begin{equation}
C\equiv \left(a+2a^2t_0^2+\frac\pi 8
e^{-2t_0^2}+2at_0^2\right)/a^{2/3} \end{equation} $t_0$ as well as
$A$ depend on $a$ only. As can be seen from Fig. \ref{C}, $C$ is
of order unity and does not vary strongly. Indeed, $C$ has a
maximum of $\Gamma\approx 1.64874$ at $1/a=8.984$ and can be
replaced by $\Gamma$ with less than 3\% error for $3<1/a<30$.

\begin{figure}
\centering \epsfig{file=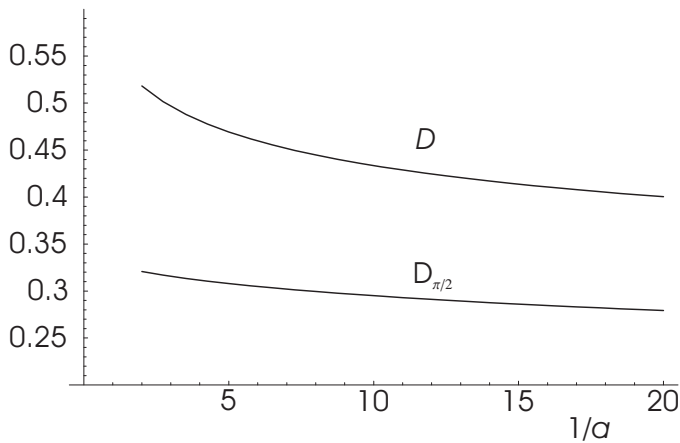,width=0.5\textwidth}
\caption{\label{Vorfaktoren} The factor $D$ (upper graph) and
$D_{\pi/2}$ (lower graph) plotted versus $1/a$.}
\end{figure}

Inserting into the expression for the integral in the saddle point
method yields
\begin{eqnarray}\label{gaussianneighbors}
g_{n\pm 2}&=&-i \frac{\bar\Om D }{\nu}\sqrt{\frac{2\sigma^2}{n}}
\exp\left\{4i\frac{(1\pm
n)}{\nu}\sqrt{\frac{2\sigma^2}{n}}\om_r t_0\right.\nonumber \\
&&\left. -4\frac{(1\pm n)^2\sigma^2\om_r^2}{\nu
a^{2/3}nC}\right\}=-g_{n\pm 2}^*.
\end{eqnarray}
The overall scaling is predominantly determined by the factor
\begin{equation}
\exp\left[-\frac{b^2}{4Ca^{2/3}}\right]=\exp\left[-4\frac{(1\pm
n)^2\sigma^2\om_r^2}{\nu a^{2/3}nC}\right].
\end{equation}
which is a strong function of $\sigma$ and, thus, $\bar\Om/\om_r$.
The factor
\begin{equation}\label{Bdef} D\equiv \frac12
\sqrt{\frac{\pi}{ Ca^{2/3}}} \sin\left[\frac\pi
4\erp(t_0)\right]e^{-at_0^2}
\end{equation}
is plotted versus $1/a$ in Fig. \ref{Vorfaktoren}.

\subsubsection{$\pi/2$ pulses}\label{gaussianpi2} $\pi/2$ pulses can be treated in
analogy by requiring \begin{equation} \frac12 \int
\Omeff(t)dt=\frac\pi4\,. \end{equation} The calculation of the
previous sections can be carried through in analogy, inserting
factors of $1/2$. For example, equation (\ref{Gaussiang}) will
have factors of $\pi/8$ in the cosine and sine functions, rather
than $\pi/4$. This means, that $g_{-n\pm2}$ will no longer be
equal to $-g_{n\mp 2}^*$, so we have to treat these cases
separately. In the following, we specialize on $g_{n\pm 2}$; the
other ones can be treated in analogy. For the evaluation of the
integral in the saddle point method, the changes can be summarized
by replacing $C$ and thus $D$ by
\begin{eqnarray} C_{\pi/2}&\equiv &
\left(a+2a^2t_0^2+\frac{\pi}{32}e^{-2t_0^2}+2at_0^2\right)/a^{2/3}\,,\nonumber \\
D_{\pi/2}&\equiv&
\frac{\pi}{2\sqrt{C_{\pi/2}n^{1/3}}}\sin\left[\frac\pi8\erp(t_0)\right]e^{-t_0^2/n}\,.
\end{eqnarray} (note that also $t_0$ will have another value).
$C_{\pi/2}$ has a maximum value of $\Gamma_{\pi/2}\approx 1.5043$
at $n=9.534$ and can be replaced by that value with little error
for $3\leq n\lesssim 30$. $D_{\pi/2}$ is also plotted in Fig.
\ref{Vorfaktoren}. It is a slowly varying function of $n$ with a
value somewhat lower than that for $\pi$ pulses. However, in
general losses for $\pi/2$ pulses will be higher, because of the
lower $\sigma$ that enters the exponential which chiefly
determines the magnitude.


\begin{thebibliography}{99}
\bibitem{remark} This condition also ensures that the argument of
$\Omeff$ changes adiabatically, and that the overall phase factor
$e^{-i\varphi}$ specified in Sec. \ref{adiabaticbragg} can be
negelected, as $\varphi \ll 1$.
\end{thebibliography}
\end{document}